\documentclass[sn-basic,Numbered]{sn-jnl}
\usepackage{graphicx}%
\usepackage{multirow}%
\usepackage{amsmath,amssymb,amsfonts}%
\usepackage{amsthm}%
\usepackage{mathrsfs}%
\usepackage[title]{appendix}%
\usepackage[table]{xcolor}
\usepackage{textcomp}%
\usepackage{manyfoot}%
\usepackage{booktabs}%
\usepackage{algorithm}%
\usepackage{algorithmicx}%
\usepackage{algpseudocode}%
\usepackage{listings}%
\usepackage{float}
\usepackage{astjnlabbrev}
\usepackage[export]{adjustbox}
\usepackage{hyperref}
\usepackage{academicons}
\usepackage{xcolor}
\usepackage{orcidlink}
\usepackage{lineno}
\usepackage{caption}

\begin{document}

\title[Article Title]{Detection of an Earth-sized exoplanet orbiting the nearby ultracool dwarf star SPECULOOS-3}



\author*[1]{\fnm{Micha\"el} \sur{Gillon}\, \orcidlink{0000-0003-1462-7739}}
\email{michael.gillon@uliege.be}

\author[2, 3]{Peter P. Pedersen\, \orcidlink{0000-0002-5220-609X}}

\author[4,5,6]{Benjamin V.\ Rackham\, \orcidlink{0000-0002-3627-1676}}

\author[7]{Georgina Dransfield\, \orcidlink{0000-0002-3937-630X}}

\author[1, 8]{Elsa Ducrot\, \orcidlink{0000-0002-7008-6888}}

\author[1,4,9]{Khalid Barkaoui\, \orcidlink{0000-0003-1464-9276}}

\author[4]{Artem Y. Burdanov\,\orcidlink{0000-0001-9892-2406}}

\author[10]{Urs Schroffenegger}

\author[11]{Yilen G\'omez Maqueo Chew\, \orcidlink{0000-0002-7486-6726}}

\author[12]{Susan M. Lederer\, \orcidlink{0000-0003-2805-8653}}

\author[9,13]{Roi Alonso\, \orcidlink{0000-0001-8462-8126}}

\author[14]{Adam J.\ Burgasser\, \orcidlink{0000-0002-6523-9536}}

\author[15]{Steve~B.~Howell\, \orcidlink{0000-0002-2532-2853}}

\author[16, 17, 9]{Norio Narita\, \orcidlink{0000-0001-8511-2981}} 

\author[4]{Julien de Wit\, \orcidlink{0000-0003-2415-2191}}

\author[10]{Brice-Olivier Demory\, \orcidlink{0000-0002-9355-5165}}

\author[2, 3]{Didier Queloz\, \orcidlink{0000-0002-3012-0316}}

\author[7]{Amaury H. M. J. Triaud\, \orcidlink{0000-0002-5510-8751}}

\author[1, 18, 19]{Laetitia Delrez\, \orcidlink{0000-0001-6108-4808}}

\author[18]{Emmanu\"el Jehin\, \orcidlink{0000-0001-8923-488X}}

\author[2]{Matthew J.\ Hooton\, \orcidlink{0000-0003-0030-332X}}

\author[1, 20]{Lionel J. Garcia\,
\orcidlink{0000-0002-4296-2246}}

\author[2]{Cl\`audia ~Jano Mu\~noz}

\author[21]{Catriona A. Murray\, \orcidlink{0000-0001-8504-5862}}

\author[22]{Francisco J. Pozuelos\, \orcidlink{0000-0003-1572-7707}}

\author[7]{Daniel Sebastian\, \orcidlink{0000-0002-2214-9258}}

\author[1]{Mathilde Timmermans\, \orcidlink{0009-0008-2214-5039}}

\author[2]{Samantha J. Thompson\, \orcidlink{0000-0002-8039-194X}}

\author[1]{Sebasti\'an Z\'u\~niga-Fern\'andez\, \orcidlink{0000-0002-9350-830X}}

\author[22, 23]{Jes\'us Aceituno\,
\orcidlink{0000-0003-0487-1105}}

\author[14, 24]{Christian Aganze\, \orcidlink{0000-0003-2094-9128}}

\author[22]{Pedro J. Amado\,}

\author[7]{Thomas Baycroft\,
\orcidlink{0000-0002-3300-3449}}

\author[25]{Zouhair Benkhaldoun\, \orcidlink{0000-0001-6285-9847}}

\author[3,26]{David Berardo\, \orcidlink{0000-0001-6298-412X}}

\author[27,28]{Emeline Bolmont\,
\orcidlink{0000-0001-5657-4503}}

\author[29,30]{Catherine~A.~Clark\, \orcidlink{0000-0002-2361-5812}}

\author[7]{Yasmin T. Davis\,
\orcidlink{0009-0000-6625-137X}}

\author[1]{Fatemeh Davoudi\,
\orcidlink{0000-0002-1787-3444}}

\author[4]{Zo\"e L. de Beurs\, \orcidlink{0000-0002-7564-6047}}

\author[31]{Jerome P. de Leon\, \orcidlink{0000-0002-6424-3410}} 

\author[32,33,34]{Masahiro Ikoma\, \orcidlink{0000-0002-5658-5971}} 

\author[31]{Kai Ikuta\, 
\orcidlink{0000-0002-5978-057X}} 

\author[35, 31]{Keisuke Isogai\, \orcidlink{0000-0002-6480-3799}} 

\author[31]{Izuru Fukuda\, \orcidlink{0000-0002-9436-2891}} 

\author[16, 9]{Akihiko Fukui\, \orcidlink{0000-0002-4909-5763}} 

\author[14,36]{Roman Gerasimov\, \orcidlink{0000-0003-0398-639X}}

\author[1, 25]{Mourad Ghachoui\, \orcidlink{0000-0003-3986-0297}}

\author[37]{Maximilian N. G\"unther\, \orcidlink{0000-0002-3164-9086}}

\author[4]{Samantha Hasler\,
\orcidlink{00000-0002-4894-193X}}

\author[31]{Yuya Hayashi\, \orcidlink{0000-0001-8877-0242}} 

\author[38]{Kevin Heng\, 
\orcidlink{0000-0003-1907-5910}}

\author[29,39]{Renyu Hu\, 
\orcidlink{0000-0003-2215-8485}}

\author[31]{Taiki Kagetani\, \orcidlink{0000-0002-5331-6637}} 

\author[31]{Yugo Kawai\, 
\orcidlink{0000-0002-0488-6297}} 

\author[40]{Kiyoe Kawauchi\, \orcidlink{0000-0003-1205-5108}} 

\author[10]{Daniel Kitzmann\, \orcidlink{0000-0003-4269-3311}}

\author[41]{Daniel D. B. Koll\, \orcidlink{0000-0002-9076-6901}}

\author[27]{Monika Lendl\, \orcidlink{0000-0001-9699-1459}} 

\author[17, 42, 33]{John H. Livingston\, \orcidlink{0000-0002-4881-3620}} 

\author[41]{Xintong Lyu\, \orcidlink{0009-0004-9766-036X}}

\author[10]{Erik A. Meier Vald\'es\,
\orcidlink{00000-0002-2160-8782}}

\author[31]{Mayuko Mori\, 
\orcidlink{0000-0003-1368-6593}} 

\author[43]{James J. McCormac\, \orcidlink{0000-0003-1631-4170}}

\author[9, 12]{Felipe Murgas\, \orcidlink{0000-0001-9087-1245}}

\author[4]{Prajwal Niraula\, \orcidlink{0000-0002-8052-3893}}

\author[9,13]{Enric Pallé\, \orcidlink{0000-0003-0987-1593}} 

\author[44]{Ilse Plauchu-Frayn\, \orcidlink{0000-0003-0987-1593}}

\author[9]{Rafael Rebolo\, \orcidlink{0000-0003-3767-7085}}

\author[44]{Laurence Sabin\, \orcidlink{0000-0003-0242-0044}}

\author[3]{Yannick Schackey}

\author[45, 94]{Nicole Schanche\, \orcidlink{0000-0002-9526-3780}}

\author[46]{Franck Selsis\, \orcidlink{0000-0001-9619-5356}}

\author[22]{Alfredo Sota\, \orcidlink{0000-0002-9404-6952}}

\author[18, 1]{Manu Stalport\, \orcidlink{0000-0003-0996-6402}}

\author[7, 47]{Matthew R.\ Standing\,
\orcidlink{0000-0002-7608-8905}}

\author[48]{Keivan G.\ Stassun\, \orcidlink{0000-0002-3481-9052}}

\author[49, 17, 42]{Motohide Tamura, \orcidlink{0000-0002-6510-0681}} 


\author[14]{Christopher A.\ Theissen\, \orcidlink{0000-0002-9807-5435}}

\author[50, 46]{Martin Turbet\, \orcidlink{0000-0003-2260-9856}}

\author[18]{Val\'erie Van Grootel\, \orcidlink{0000-0003-2144-4316}}

\author[22]{Roberto Varas\,
\orcidlink{0000-0002-6946-0342}}

\author[31]{Noriharu Watanabe\, \orcidlink{0000-0002-7522-8195}} 

\author[10]{Francis Zong Lang\,
\orcidlink{00009-0005-3162-3694}}

\affil*[1]{\orgdiv{Astrobiology Research Unit}, \orgname{Universit\'e de Li\`ege}, \orgaddress{\street{All\'ee du 6 ao\^ut 19}, \city{Li\`ege}, \postcode{4000}, \country{Belgium}}}

\affil[2]{Cavendish Laboratory, JJ Thomson Avenue, Cambridge CB3 0HE, UK}

\affil[3]{Department of Physics, ETH Zurich, Wolfgang-Pauli-Strasse 2, CH-8093 Zurich, Switzerland}

\affil[4]{Department of Earth, Atmospheric and Planetary Science, Massachusetts Institute of Technology, 77 Massachusetts Avenue, Cambridge, MA 02139, USA}

\affil[5]{Kavli Institute for Astrophysics and Space Research, Massachusetts Institute of Technology, Cambridge, MA, USA}

\affil[6]{51 Pegasi b Fellow}

\affil[7]{School of Physics \& Astronomy, University of Birmingham, Edgbaston, Birmingham B15 2TT, UK}

\affil[8]{LESIA, Observatoire de Paris, CNRS, Universit\'e Paris Diderot, Universit\'e Pierre et Marie Curie, 5 place Jules Janssen, 92190 Meudon, France}

\affil[9]{Instituto de Astrof\'isica de Canarias (IAC), Calle V\'ia L\'actea s/n, 38200, La Laguna, Tenerife, Spain}

\affil[10]{Center for Space and Habitability, University of Bern, Gesellschaftsstrasse 6, CH-3012 Bern, Switzerland}

\affil[11]{Instituto de Astronom\'ia, Universidad Nacional Aut\'onoma de M\'exico, Ciudad Universitaria, Ciudad de M\'exico 04510, M\'exico}

\affil[12]{NASA Johnson Space Center, 2101 NASA Parkway, Houston, Texas, 77058, USA}

\affil[13]{Departamento de Astrof\'\i sica, Universidad de La Laguna, Astrofísico Francisco S\'anchez s/n, 38206 La Laguna, Tenerife, Spain}

\affil[14]{Department of Astronomy \& Astrophysics, University of California San Diego, La Jolla, CA 92093, USA}

\affil[15]{NASA Ames Research Center, Moffett Field, CA 94035, USA}

\affil[16]{Komaba Institute for Science, The University of Tokyo, 3-8-1 Komaba,
Meguro, Tokyo 153-8902, Japan}

\affil[17]{Astrobiology Center, 2-21-1 Osawa, Mitaka, Tokyo 181-8588, Japan}

\affil[18]{Space Sciences, Technologies and Astrophysics Research (STAR) Institute,
Universit\'e de Li\`ege, All\'ee du 6 Ao\^ut 19C, B-4000 Li\`ege, Belgium}

\affil[19]{Institute of Astronomy, KU Leuven, Celestijnenlaan 200D, 3001 Leuven, Belgium}

\affil[20]{Center for Computational Astrophysics, Flatiron Institute, 162 Fifth Avenue, New York, New York 10010, USA}

\affil[21]{Department of Astrophysical and Planetary Sciences, University of Colorado Boulder, Boulder, CO 80309, USA}

\affil[22]{Instituto de Astrof\'isica de Andaluc\'ia (IAA-CSIC), Glorieta de la Astronom\'ia s/n, E-18008 Granada, Spain}

\affil[23]{Centro Astron\'omico Hispano en Andalucia, Sierra de los Filabres sn , 04550 G\'ergal Almer\'ia, Spain}

\affil[24]{Department of Physics, Stanford University, Stanford, CA 94305, USA}

\affil[25]{Oukaimeden Observatory, High Energy Physics and Astrophysics Laboratory, Faculty of Sciences Semlalia, Cadi Ayyad University, Marrakech, Morocco}

\affil[26]{Department of Physics and Kavli Institute for Astrophysics and Space Research, Massachusetts Institute of Technology, Cambridge, MA 02139, USA}

\affil[27]{Département d'astronomie de l’Université de Genève, Chemin Pegasi 51, 1290 Sauverny, Switzerland}

\affil[28]{Centre sur la Vie dans l'Univers, Université de Genève, Switzerland}

\affil[29]{Jet Propulsion Laboratory, California Institute of Technology, Pasadena, CA 91109 USA}

\affil[30]{NASA Exoplanet Science Institute, IPAC, California Institute of Technology, Pasadena, CA 91125 USA}

\affil[31]{Department of Multi-Disciplinary Sciences, Graduate School of Arts and Sciences, The University of Tokyo, 3-8-1 Komaba, Meguro, Tokyo 153-8902, Japan}

\affil[32]{Division of Science, National Astronomical Observatory of Japan, 2-21-1 Osawa, Mitaka, Tokyo 181-8588, Japan}

\affil[33]{Astronomical Science Program, Graduate University for Advanced Studies, SOKENDAI, 2-21-1, Osawa, Mitaka, Tokyo, 181-8588, Japan}

\affil[34]{Department of Earth and Planetary Science, The University of Tokyo, 7-3-1 Hongo, Bunkyo-ku,
Tokyo 113-0033, Japan}

\affil[35]{Okayama Observatory, Kyoto University, 3037-5 Honjo, Kamogatacho, Asakuchi, Okayama 719-0232, Japan}

\affil[36]{Department of Physics \& Astronomy, University of Notre Dame, Notre Dame, IN 46556, USA}

\affil[37]{European Space Agency (ESA), European Space Research and Technology Centre (ESTEC), Keplerlaan 1, NL-2201 AZ Noordwijk, the Netherlands}

\affil[38]{Faculty of Physics, Ludwig Maximilian University, Scheinerstrasse 1, D-81679, Munich, Bavaria, Germany}

\affil[39]{Division of Geological and Planetary Sciences, California Institute of Technology, Pasadena, CA 91125, USA}

\affil[40]{Department of Physical Sciences, Ritsumeikan University, Kusatsu, Shiga 525-8577, Japan}

\affil[41]{Department of Atmospheric and Oceanic Sciences, Peking University,
Beijing, China}

\affil[42]{National Astronomical Observatory of Japan, 2-21-1 Osawa, Mitaka, Tokyo 181-8588, Japan}

\affil[43]{Department of Physics, University of Warwick, Gibbet Hill Road, Coventry CV4 7AL, UK}

\affil[44]{Instituto de Astronom\'ia, Universidad Nacional Aut\'onoma de M\'exico, Unidad Acad\'emica en Ensenada, 22860 Ensenada, B.C., M\'exico}

\affil[45]{NASA Goddard Space Flight Center, 8800 Greenbelt Rd, Greenbelt, MD 20771, USA}

\affil[46]{Laboratoire d'astrophysique de Bordeaux, Univ. Bordeaux, CNRS, B18N, allée Geoffroy Saint-Hilaire, 33615 Pessac, France}

\affil[47]{European Space Agency (ESA), European Space Astronomy Centre (ESAC), Camino Bajo del Castillo s/n, 28692 Villanueva de la Ca\~{n}ada, Madrid, Spain}

\affil[48]{Department of Physics and Astronomy, Vanderbilt University, Nashville, TN 37235, USA}

\affil[49]{Department of Astronomy, University of Tokyo, 7-3-1 Hongo, Bunkyo-ku,
Tokyo 113-0033, Japan}

\affil[50]{Laboratoire de Météorologie Dynamique/IPSL, CNRS, Sorbonne Université, Ecole Normale Supérieure, Université PSL, Ecole Polytechnique, Institut Polytechnique de Paris, 75005 Paris, France}

\affil[94]{Department of Astronomy, University of Maryland, College Park, MD  20742, USA}



\abstract{Located at the bottom of the main-sequence, ultracool dwarf stars are widespread in the solar neighborhood. Nevertheless, their extremely low luminosity has left their planetary population largely unexplored, and only one of them, TRAPPIST-1, has so far been found to host a transiting planetary system. In this context, we present the SPECULOOS project's detection of an Earth-sized planet in a 17-hour orbit around an ultracool dwarf of M6.5 spectral type located 16.8 parsecs away. The planet's high irradiation (16 times that of Earth) combined with the infrared luminosity and Jupiter-like size of its host star make it one of the most promising rocky exoplanets for detailed emission spectroscopy characterization with JWST. Indeed, our sensitivity study shows that just 10 secondary eclipse observations with the Mid-InfraRed Instrument Low-Resolution Spectrometer on board JWST should bring strong constraints on its atmospheric composition and/or surface mineralogy. }




\maketitle

\section{Main}\label{sec1}

\subsection{Introduction}

At the end of the main sequence, ultracool dwarf stars (UDS) \cite{Kirkpatrick1997} have spectral types between M6.5 and L2, masses between 0.07 and 0.1 solar masses, sizes similar to Jupiter, and effective temperatures between 2200 and 2850\,K \cite{Dieterich2014}. The SPECULOOS (Search for Planets EClipsing ULtra-cOOl Stars) project \cite{Gillon2018, Burdanov2018, Delrez2018} aims to perform a volume-limited (40\,pc) transit search on $\sim$1650 very-low-mass stars and brown dwarfs, including $\sim$900 UDS \cite{Sebastian2021}. It is based on a network of six robotic 1-m-aperture telescopes: the four telescopes of the SPECULOOS-South Observatory (SSO) in Chile \cite{Jehin2018}, Artemis, the first telescope of the SPECULOOS-North Observatory (SNO) in Tenerife \cite{Burdanov2022}, and the SAINT-EX telescope in San Pedro M\'artir Observatory in Mexico \cite{Demory2020}.
SPECULOOS achieved its first main result in 2015 with the discovery, by its prototype on the robotic 60\,cm telescope TRAPPIST \cite{Gillon2013}, of a system composed of seven Earth-sized planets in close orbits around an M8-type UDS at 12 pc \cite{Gillon2016, Gillon2017}. More recently, SPECULOOS discovered a super-Earth orbiting in the habitable zone of LP\,890-9 (also known as SPECULOOS-2, TOI-4306), an M6-type dwarf star at 32 pc \cite{Delrez2022}. 

SPECULOOS-3 (aka LSPM J2049+3336) is an isolated UDS of spectral type M$6.5 \pm 0.5$ located 16.75 pc away. It is part of the $\sim$365 SPECULOOS Program 1 targets that are small and nearby enough to make the detailed atmospheric characterisation of a temperate Earth-sized planet possible with JWST \cite{Sebastian2021}. It was observed by SAINT-EX for five nights in 2021, and by SNO-Artemis for three nights in 2022. The visual inspection of the 2021 and 2022 light curves showed some transit-like structures that motivated future intensive monitoring of the star. It was thus re-observed by SNO-Artemis from July to September 2023. In total, 18 transit-like structures were observed by SNO-Artemis in 2023 and could be related to a period of $\sim$0.72\,d. Furthermore, the other transit-like structures in the 2021 (SAINT-EX) and 2022 (SNO-Artemis) light curves corresponded to the identified transit ephemeris (Fig. 1). Follow-up photometric observations were also made to confirm the transit and its achromatic nature using the SSO 1-m telescopes, the TRAPPIST-North 0.6-m telescope at Ouka\"imeden Observatory in Morocco, the 1.5-m telescope at Sierra Nevada Observatory in Spain, the MuSCAT3 instrument on the 2-m Faulkes North telescope in Hawaii, the 3.8-m UK Infrared Telescope (UKIRT) in Hawaii, and the HiPERCAM instrument on the 10-m GTC telescope at La Palma (Fig. 2).

 New spectroscopic observations with the Kast Double Spectrograph  mounted on the 3-m Shane telescope at Lick Observatory (Fig. 3, top), the SpeX spectrograph mounted on the 3-m NASA InfraRed Telescope Facility (IRTF) telescope in Hawaii  (Fig. 3, bottom), and the CARMENES spectrograph mounted on the 3.5-m telescope at Calar Alto Observatory, as well as high-resolution images obtained with the `Alopeke instrument on the Gemini-North telescope in Hawaii, were also gathered. Combined with archival absolute magnitudes and the Gaia EDR3 parallax, some of the new spectra, once flux-calibrated, enabled the luminosity of the star to be measured as $0.000835 \pm 0.000019\,\rm L_\odot$, following the procedure described in \cite{Stassun:2018} and references therein.

\subsection{Results}

 Two astrophysical cases had to be excluded to confirm the planetary nature of the candidate. The first case is that of a background eclipsing binary blended with the target's images in the data. Thanks to the high proper motion of the target (0.46 as/yr), archival images directly discarded this hypothesis by confirming that no background source of significant brightness is located behind its current position (Extended Data Fig. 1). The second case is that of a bound eclipsing binary companion. Several factors argue against this hypothesis when considered individually, and fully discard it when considered globally: (1) the stability of the star's radial velocity, at the 50\,m\,s$^{-1}$ level over a 1-yr period, as inferred from APOGEE spectra \cite{Birky2020}, and the discarding of a radial velocity slope larger than 5\,m\,s$^{-1}$ per day over the two months period covered by observations of the high-resolution spectrograph CARMENES; (2) the non-detection of a companion object in the GEMINI-North high-resolution images (Extended Data Fig. 2); (3) the non-detection of a secondary spectrum in the CARMENES, Kast and SpeX spectra (Fig. 3); (4) the fact that the global spectral energy distribution (SED) of the target is well-fitted by the spectral model of an isolated $\sim$M6.5 dwarf (Extended Data Fig. 3); (5) the excellent agreement between the stellar density inferred from the transits ($55.1_{-3.7}^{+2.0} \,\rho_\odot$) and that deduced from the basic parameters of the star ($55 \pm 15 \,\rho_\odot$); and (6) the achromatic nature of the transit depth from $\sim$470 nm ($g'$ filter) to $\sim$2.2 $\mu$m (Fig. 2). 
 
 The age of the system is constrained to be 6.6$^{+1.8}_{-2.4}$~Gyr from its kinematics (see Methods), and the star's luminosity, mass and radius are 0.084$\pm$0.002\%, 10.1$\pm$0.2\% and 12.3$\pm$0.2\% of those of the Sun, respectively. Just slightly larger than TRAPPIST-1,  SPECULOOS-3 is the second-smallest main-sequence star found to host a transiting planet (Fig. 4, left). The small size of the host star---only slightly larger than Jupiter---translates into an Earth-like radius for the transiting planet, as deduced from its $\sim$0.5\% transit depth. Table 1 presents the physical properties of the system, as derived through a global Bayesian analysis of the transit photometry, including the a priori knowledge of its stellar properties, with an adaptive Markov chain Monte Carlo (MCMC) code (see Methods).

The planet is very similar in size to the Earth: 0.977 $\pm$ 0.022~$R_\oplus$. Its equilibrium temperature is 553 $\pm$ 8K, assuming a null Bond albedo and a full heat redistribution. Its mass, and thus its composition, remains unconstrained by our observations thus far. Nevertheless, several factors strongly suggest a rocky composition. From a theoretical point of view, the intense extreme-ultraviolet (1–1,000 \AA) emission of low-mass stars during their early lives \cite{Stelzer2013} makes it unlikely that such a small planet on such a short orbit could have maintained a substantial envelope of hydrogen \cite{Lopez2012, Owen2016}. From an empirical point of view, SPECULOOS-3\,b falls well within the rocky side of the {\it radius gap}, i.e. the paucity of planets with radii between 1.5 and 2 $R_{\oplus}$ attributed to photo-evaporative processes \cite{Owen2013, Owen2017, Fulton2017, Luque2022, Petigura2022}. We also note that the list of all currently known Earth-sized planets in the NASA exoplanet archive \cite{Akeson2013} have masses implying rocky compositions.

Ultimately, measuring the mass of SPECULOOS-3\,b is essential to determine whether it is indeed rocky and to further constrain its composition. By our estimates, a reasonable observing programme ($<5$ nights) that would be able to detect the radial velocity signal of the planet's Doppler reflex motion could be accomplished using state-of-the-art spectrographs (see Methods). Such observations should be able to differentiate between Earth-like, iron-poor, and water-rich compositions \cite{Luque2022}.\\

\subsection{Discussion}

TRAPPIST-1\,b is the closest analog to SPECULOOS-3\,b in terms of host star size (0.123 $vs$ 0.119 $R_\odot$), planet size (0.98 $vs$ 1.12 $R_\oplus$) and equilibrium temperature (553 $vs$ 400K)
(see Fig. 4). Nevertheless, the two systems appear to be very different: TRAPPIST-1 hosts a resonant system of seven short-orbit rocky planets, while an intensive photometric monitoring campaign of SPECULOOS-3 with SNO and MuSCAT-3 failed to detect any outer transiting planets over a 10 days period range (see Methods). This indicates that both systems had different formation/evolution histories. As outlined by recent works, resonant systems like TRAPPIST-1 are rare because they are prone to disruption by dynamical instabilities \cite{Izidoro2017, Izidoro2021, Goldberg2022}. SPECULOOS-3\,b could thus be the outcome of such a disruption. Outer planets may still exist, but on much longer and/or mutually inclined orbits.  Nevertheless, we cannot exclude the possibility that outer well-aligned planets do exist, but that they are just too small to be detected with current observations (see Methods).  

With current irradiation 16 times larger than Earth's, which should have been much larger during the $\sim$800 Myr pre-main sequence phase of the star, the possibility for it to have kept a substantial secondary atmosphere is slim. Nevertheless, the possibility exists that a volatile-rich initial composition could have sustained a steady-state secondary atmosphere despite this adverse environment \cite{Forget2014, Zahnle2017, Grenfell2020}. Compared to the more temperate TRAPPIST-1 planets, SPECULOOS-3\,b has the advantage that transit transmission spectroscopy \cite{Lim2023} or emission or phase curve photometry \cite{Greene2023Natur, Zieba2023} are not the only methods available to assess the presence of an atmosphere. Indeed, the planet is hot enough and its host star small and infrared-bright enough to make it possible to measure its dayside emission spectrum with the Mid-InfraRed Low-Resolution Spectrometer (MIRI/LRS) aboard JWST. Fig. 4 (right) shows how SPECULOOS-3\,b compares in terms of potential for emission spectroscopy with other known transiting terrestrial planets ($R_{\rm{p}}<1.6 R_\oplus$) that are cool enough ($T_{\rm{eq}}<$ 880 K) to have a dayside made of solid rock. Planets with $T_{\rm{eq}}>$ 880 K are expected to have molten (lava) surfaces and no atmospheres, except maybe for vaporized rocks \cite{2019ApJ...886..141M}. The potential of the planets for emission spectroscopy is quantified using the Emission Spectroscopy Metric (ESM) \cite{kempton_framework_2018}, which is proportional to the expected S/N of a JWST secondary eclipse detection at mid-IR wavelengths. With an ESM value of 7.8, SPECULOOS-3\,b is currently one of the smallest planets above the recommended threshold of ESM=7.5 \cite{kempton_framework_2018} that identifies the top targets for emission spectroscopy with JWST.

Such observations could not only reveal the presence of an atmosphere while avoiding the critical problem of stellar contamination inherent to transit transmission spectroscopy \cite{2018ApJ...853..122R} (as observed for TRAPPIST-1\,b \cite{Lim2023} or GJ 1132\,b \cite{2023arXiv231010711M}, see details in Methods), but could also constrain the mineralogical surface of this Earth-sized exoplanet, if airless. 
We explored the possibility to reveal the nature of the planet with the JWST in emission. 
We simulated emission spectroscopy observations with the JWST MIRI/LRS mode using \texttt{PandExo} (see Methods). We then modeled the emission spectra of several plausible atmospheric scenarios and bare rock surfaces (see Methods). These models include CO$_2$-dominated or H$_2$O-dominated atmospheres, as well as surfaces resulting from various geological configurations, either primary crusts from the solidification of a magma ocean (ultramafic and feldspathic), secondary crusts produced by volcanic eruptions (basaltic), or tertiary crusts produced by plate tectonics (granitoid). We find that in only 10 occultations observed with MIRI/LRS we can reach the required precision to assess the presence of an atmosphere and distinguish between the most plausible atmospheric scenarios or, if the planet is airless, distinguish between 50\% of competing surface models at 4 $\sigma$ (Fig. 5, Extended Data Tables 1 and 2). Finally, if the planet is airless, emission spectroscopy can also be used to reveal the planet's geological history by constraining the surface emissivity spectrum and indirectly its albedo. Indeed, old bare-rock planets are expected to be much darker than the ones with fresh geologic surfaces due to the effect of space weathering (as observed on Mercury or the Moon \cite{2001JGR...10610039H}). Space weathering timescales are short ($\simeq$ 100 years) for close-in planets around active stars like SPECULOOS-3\,b, such that any detection of a high-albedo dayside would require that the planet's surface to be geologically very young and indicate the presence of active volcanism or tectonic overturn (Extended Data Fig. 4).

These results motivate a dedicated study of SPECULOOS-3\,b through emission spectroscopy with JWST, and encourage us to pursue the search and detailed study of the still poorly understood terrestrial planets in orbit around the ubiquitous ultracool dwarf stars. 

\noindent

\newpage 
\section{Methods}\label{sec11}

\noindent\textbf{Discovery SPECULOOS photometry. } The first SPECULOOS observations of the target were acquired in 2019 and 2021 with the SAINT-EX 1m telescope \cite{Demory2020} in the $I+z$ filter. Two of the 2021 light curves included a transit of SPECULOOS-3\,b. They were noticed, but the star was not ranked yet as a high-priority target because of the high level of correlated noise in these two light curves due to bad weather conditions. In July 2022, the SPECULOOS-North Observatory Artemis (SNO-Artemis) 1-m telescope \cite{Burdanov2022} observed the target for three nights, also in the $I+z$ filter. One of the three resulting light curves (29 Jul 2022) showed a clear transit-like signature. An intensive monitoring of the star was then initiated from July 2023, first with SNO-Artemis. Several clear transit-like structures showed up in subsequent light curves. They could be related to a period of $\sim$0.719d, and their shape (duration, depth) was consistent with an Earth-sized planet transiting the star (Fig. 1). SNO-Artemis pursued its monitoring of the star up to the end of Sep 2023, resulting in 37 light curves of 2 hrs or more.  Several transit windows were also observed in Aug and Sep 2023 with some of or all the four 1-m telescopes of the SPECULOOS-South Observatory \cite{Jehin2018} at ESO Cerro Paranal Observatory in Chile. Some of these observations were taken in the $I+z$ filter, others in $i'$ and $z'$ filters to assess the chromaticity of the transit. 
In total, SNO-Artemis observed 18 transits of SPECULOOS-3\,b, and SSO telescopes observed 5 of them (see Supplementary Information Fig. 1). \\
\\ 
\textbf{Host star properties.}  SPECULOOS-3 = 2MASS J20492745+3336512 = LSPM J2049+3336 was identified as a high-proper-motion stellar object in 2005 \cite{Lepine2005}, and as a nearby late-type M-dwarf in 2014 \cite{Dittmann2014}. Spectra taken by the SDSS-3 APOGEE survey led to the first effective temperature estimate of 2765-2800\,K and an upper limit of $8$\, km\,s$^{-1}$ for its  $v\sin{i_\star}$ \cite{Gilhool2018}. 
The three barycentric radial velocities measured from APOGEE spectra (two in 2013, one in 2014) were stable at the 50 m\,s$^{-1}$ level, supporting the hypothesis of an isolated star \cite{Jonsson2020}. Based on photometric data and the star's Gaia DR2 parallax (59.733 $\pm$ 0.088 mas, \cite{GaiaDR2}), \cite{Scholz2020} estimated a spectral type of M$6.5 \pm 1.5$. 

We observed SPECULOOS-3 twice with the SpeX near-infrared spectrograph \citep{Rayner2003} on the 3.2-m NASA Infrared Telescope Facility (IRTF). On 30 Aug 2021 (UT), we collected a medium-resolution spectrum ($\lambda / \Delta \lambda {\sim} 2000$) 
using the short-wavelength cross-dispersed (SXD) mode, covering 0.80--2.42\,$\mu$m. 
On 11 Aug 2023 (UT), we collected a low-resolution spectrum ($\lambda / \Delta \lambda {\sim} 120$) using the prism-dispersed mode, covering 0.70--2.52\,$\mu$m. 
All data were reduced using Spextool v4.1 \citep{Cushing2004}. The final spectra are shown in Fig. 3 (bottom panel). The SXD and prism spectra have median signal-to-noise ratios (SNR) of 55 and 250 per pixel, respectively.
To assign an infrared spectral type, we used the SpeX Prism Library Analysis Toolkit \citep[SPLAT, ][]{splat} to compare the SXD spectrum to 
spectral standards in the IRTF Spectral Library \citep{Cushing2005,Rayner2009}, finding
a best match to the M6 dwarf Wolf~359 (Fig. 3, bottom panel). 
We also compared the prism data to low-resolution standards defined in \cite{Kirkpatrick2010}, and find a best-match in the 0.8--1.3~$\mu$m range to the M7 standard VB~8.
We therefore adopt an infrared spectral type of M6.5 $\pm$ 0.5 for SPECULOOS-3.
Using the relation of ref.\,\cite{Mann_met} and following the approach of ref.\,\citep{Delrez2022}, we estimate an iron abundance of $\mathrm{[Fe/H]} = +0.08 \pm 0.11$.

We obtained a red optical spectrum of SPECULOOS-3 with the Kast Double Spectrograph \citep{kastspectrograph} on the 3-m Shane Telescope at Lick Observatory on 29 Sep 2022 (UT). We used the red channel with 600/7500 grating and $1.5''$-wide slide to obtain 6000--9500~{\AA} spectra at an average resolution of $\lambda/\Delta\lambda$ = 1900. 
Data were reduced using the \texttt{kastredux} package available at \url{https://github.com/aburgasser/kastredux}. The final spectrum (Fig. 3, top panel) has a SNR of 90 at 8350\,{\AA}.

We determined an optical classification by comparing it to late-M dwarf spectral templates from \cite{2017ApJS..230...16K}, and found the M7 standard provided the best fit. Index-based classifications \cite{1997AJ....113..806G,1999AJ....118.2466M,2003AJ....125.1598L} indicate a classification closer to M6. We therefore adopt a mean optical classification of M6.5$\pm$0.5, consistent with our infrared data. We measure the metallicity statistic $\zeta$ = 1.026$\pm$0.005 \citep{2013AJ....145..102L}, consistent with [Fe/H] = +0.04$\pm$0.20 using the calibration of \cite{2013AJ....145...52M}, also consistent with our infrared data.

We performed an analysis of the broadband spectral energy distribution (SED) of the star together with the Gaia DR3 parallax \citep[with no systematic offset applied; see, e.g.,][]{StassunTorres:2021}, in order to determine an empirical measurement of the stellar radius, following the procedures described in \cite{Stassun:2016,Stassun:2017,Stassun:2018}. We pulled  the $JHK_S$ magnitudes from 2MASS, the W1--W3 magnitudes from  WISE, the $G_{\rm BP}$ and $G_{\rm RP}$ magnitudes from  Gaia, and the $gzy$ magnitudes from  Pan-STARRS. Together, the available photometry spans the full stellar SED over the wavelength range 0.4--10~$\mu$m (see Extended Data Fig. 3). We also utilized for comparison the absolute flux-calibrated Kast and SpeX spectrophotometry (see above). We performed a fit using PHOENIX stellar atmosphere models \cite{Husser:2013}, with the free parameters being the effective temperature ($T_{\rm eff}$) and metallicity ([Fe/H]), taking the extinction $A_V \equiv 0$ due to the proximity of the system to Earth. The resulting fit (Extended Data Fig. 3) has a best-fit $T_{\rm eff} = 2680 \pm 60$~K and [Fe/H] = $-0.15 \pm 0.25$, with a reduced $\chi^2$ of 2.1. Both values are consistent with the adopted values inferred as described in the next section. Integrating the model SED gives the bolometric flux at Earth, $F_{\rm bol} = 9.54 \pm 0.22 \times 10^{-11}$ erg~s$^{-1}$~cm$^{-2}$. Taking the $F_{\rm bol}$ and the {\it Gaia\/} parallax gives directly the bolometric luminosity, $L_{\rm bol} = (8.35 \pm 0.19) \times 10^{-4}$~L$_\odot$, which with the Stefan-Boltzmann relation implies a stellar radius $R_\star = 0.1342 \pm 0.0062$~R$_\odot$. We note that this radius together with the spectroscopically-derived $v\sin i$ upper limit from APOGEE ($8$\, km\,s$^{-1}$) implies a projected rotation period of 
$P_{\rm rot} / \sin i > 0.85$~d. \\

We adopted the weighted average of the Kast (visible) and SpeX (near-IR) measurements for the star's metallicity: [Fe/H] = $0.07 \pm 0.10$. Using the  Gaia EDR3 parallax and the 2MASS $H$-magnitude of the star, we estimated the effective temperature of SPECULOOS-3 to be $T_{\rm eff} = 2829 \pm 30$\,K using the empirical relationship $T_{\rm eff}$(M$_H$) of \cite{Filipazzo2015}, and accounting for its internal error of 29\,K. This $T_{\rm eff}$ value is in excellent agreement with the one inferred by \cite{Gilhool2018} and \cite{Birky2020}, but in tension with the one inferred from our SED fitting ($2680 \pm 60$~K). In this context, we decided to adopt (as prior in our global analysis, see below) the weighted average of the two estimates and  an error bar large enough to encompass the value of 2680~K from the SED-fitting, resulting in $T_{\rm eff} = 2800 \pm 120$~K. We then used the Stefan-Boltzmann law to compute the stellar radius  as $R_\star = 0.123 \pm 0.011 \rm R_\odot$ from our adopted $T_{\rm eff}$ and measured $L_{\rm bol}$. Finally, we estimated the mass of the star to be $M_\star = 0.1017 \pm 0.0024 \rm M_{\odot}$ from its 2MASS $K$-magnitude and its Gaia EDR3 parallax, using the relationship of \cite{Mann2019} and accounting for its internal error of 0.0023 $M_\odot$. 

Using the \texttt{PyAstronomy/gal\_uvw} routine \cite{PyA2019} based on the algorithm of \cite{Johnson1987}, with as input the Gaia EDR3 coordinates and proper motions of the star \cite{GaiaEDR3} and its APOGEE radial velocity \cite{Birky2020}, we computed the following galactic velocities in the Local Standard of Rest (LSR) frame for  SPECULOOS-3: $U=47.16 \pm 0.29$ km\,s$^{-1}$, $V=21.34 \pm 0.43$ km\,s$^{-1}$, and $W=-3.68 \pm 0.26$ km\,s$^{-1}$. These values were computed using the LSR correction of \cite{LSR2011}, U V W$_\odot$ = [+8.50,+13.38,+6.49]. These velocities would place the star, statistically speaking, in the thin disk of our galaxy, suggesting an age lower than $\sim$8 Gyr \cite{Li2017}.
A more careful analysis comparing the kinematics of SPECULOOS-3 to nearby co-moving stars (within 5~km\,s$^{-1}$ in $UVW$) with a similar metallicity (within 0.2~dex) using data from GALAH DR3 \cite{2021MNRAS.506..150B} yields an uncertainty-weighted average age of 6.6$^{+1.8}_{-2.4}$~Gyr.

We also ran stellar evolution modeling using structure models presented in \cite{Fernandes2019}. We used as constraints the luminosity and metallicity as derived above. We obtained a stellar mass ($M_\star=0.1001\pm0.0015 \rm M_{\odot}$), radius ($R_\star=0.125\pm0.002  \rm R_{\odot}$) and  effective temperature ($T_{\rm eff}=2780\pm 30$ K) in excellent agreement with the estimates obtained above. Finally, the pre-main sequence phase for a 0.10 $M_{\odot}$ star lasts about 800 Myr, suggesting a higher age for SPECULOOS-3.\\

\noindent\textbf{TESS photometry.} We downloaded the TESS two-minute-cadence PDCSAP Sector 41 and 55 light curve using \texttt{lightkurve} \citep{lightkurve}. The analysis of the TESS photometry of SPECULOOS-3 reveals its flaring nature, leads to a marginal detection of the transit of SPECULOOS-3\,b, and does not result in an unambiguous determination of its rotation period (see Supplementary Information Fig. 2).\\

\noindent\textbf{MuSCAT2 photometry.} A transit of SPECULOOS-3b was observed on Aug 12 2023 with the MuSCAT2 multi-imager instrument \cite{Narita2019} at the Telescopio Carlos S\'anchez (TCS) located at the Teide Observatory (Spain).  Unfortunately, the night was cloudy and the resulting photometry was too noisy to be of any scientific use.\\

\noindent\textbf{MuSCAT3 photometry.} We used the 2.0-m Faulkes Telescope North at Haleakala Observatory located in Hawaii to observe SPECULOOS-3. The telescope is equipped with the MuSCAT3 multi-band imager \cite{Narita_2020SPIE11447E}. The campaign was started on the single night of Sept 3 2023, and then from Sept 15 2023 to Sept 26 2023. Its main goal was to search for additional transiting planets in the system. During this campaign, we observed two full transits of SPECULOOS-3\,b on UT September 17 and 19 2023 in the Sloan-$g'$, -$r'$, -$i'$ and $z_s$ filters. All datasets were calibrated using the standard LCOGT {\tt BANZAI} pipeline \cite{McCully_2018SPIE10707E}, and photometric measurements were extracted in an uncontaminated target aperture using the {\tt PROSE} pipeline \citep{2022_prose}.\\

\noindent\textbf{TRAPPIST-North photometry.}
TRAPPIST-North (TRAnsiting Planets and PlanetesImals Small Telescope) is a 60-cm robotic telescope located at Oukaimeden Observatory in Morocco \cite{Gillon2011_TS,Jehin2011,Barkaoui2019_TN}. It observed a total of 6 transits of SPECULOOS-3\,b on August 2, 4, 5, 17, 28 and 30 2023 in the $I+z$ and Sloan-$z'$ filters. When combined, the resulting light curves confirmed the transit, but we did not include them in our global analysis as their precision was substantially lower than the one obtained with the other telescopes. \\

\noindent\textbf{T150 photometry.} We conducted a full-transit observation of SPECULOOS-3\,b on 20 September 2023 using the T150 at the Sierra Nevada Observatory in Granada (Spain). The T150 is a 150-cm Ritchey-Chr\'etien telescope equipped with a thermoelectrically cooled 2K$\times$2K Andor iKon-L BEX2DD CCD camera with a field of view of $7.9'\times7.9'$ and pixel scale of 0.232". We used the Johnson-Cousin $I$ filter with an exposure time of 60\,s. The photometric data were extracted using the {\tt AstroImageJ} package \citep{collins2017}.\\

\noindent\textbf{GTC/HiPERCAM photometry.} A full transit of SPECULOOS-3\,b was observed simultaneously (DDT program GTC2023-216) in five ‘Super’-SDSS filters $u_sg_sr_si_sz_s$ on September 17 2023 with the {\tt HiPERCAM} (High PERformance CAMera \cite{Dhillon_Hipercam}) instrument, mounted on the 10.4m Gran Telescopio Canarias (GTC) located in the Roque de los Muchachos Observatory in the island of La Palma.  The observations covered about 1.94~h centered around a predicted transit. We reduced the data using the {\tt HiPERCAM} pipeline, which allows to perform aperture photometry with either fixed apertures, or scaled to the full-width and half maximum (FWHM) of the individual frames. The setup that provided the most precise photometry consisted on fixed aperture sizes for the $u_s$ and $z_s$ bands, and scaled apertures for the rest. A reference star was constructed by summing the fluxes from four nearby stars in the $u_s$, $g_s$, and $r_s$ bands, and from one nearby star in the $i_s$ and $z_s$ bands.
The final light curves show a clear detection of the transit in all but the $u_s$ band, in which the target is estimated to have a magnitude of $\sim$22.7, and thus the photon noise dominates over the expected transit depth. Additionally, around 6~min after the egress of the transit, a flare is detected in the three bluest bands, with a duration of about 5~min, and peak amplitudes of 3\%, 9\%, and 140\% in the $r_s$, $g_s$, and $u_s$ bands, respectively.\\

\noindent\textbf{UKIRT photometry.}  On 27 and 30 Aug 2023, two transits of SPECULOOS-3\,b were observed in the $Ks$-filter (mean wavelength = 2.2 $\mu$m, width = 0.34 $\mu$m) with the WFCAM camera on the UKIRT 4-m telescope on Maunakea, Hawaii.  Our data analysis started from the calibrated images obtained with the WFCAM detector \#3, which included SPECULOOS-3 and dozens of stars of similar magnitude in its $13.65' \times 13.65'$ field of view.  For the first run, the full-width at half maximum (FWHM) of the target's point-spread function (PSF) ranged from about 1.5 to 2.75 pixels (WFCAM pixel scale = $0.4''$), with a median value of 1.92 $\pm$ 0.25 pixels. The PSF was thus poorly sampled in an important part of the images.
Despite the active guiding, the x- and y-positions of the star drifted $\sim 1$ pixel over the run. The fluxes of the target and comparison stars were measured in the images with \texttt{IRAF/DAOPHOT} \cite{Stetson1987}. The resulting differential light curves showed substantial correlated noise that we attribute to the FWHM variations, the drift of the stars on the detector, and the poor sampling of the PSF. For the second run, the FWHM of the target's PSF ranged from roughly 2.5 to 4.5 pixels, with a median value of 2.69 $\pm$ 0.53 pixels. The PSF was thus better sampled than for the first run, resulting in more precise photometry. Here too, the x- and y-positions of the target drifted $\sim 1$ pixel along the run. \\

\noindent\textbf{CARMENES high-resolution spectroscopy.} We observed SPECULOOS-3 with the CARMENES instrument \citep{quirrenbach2020} installed at the 3.5-m telescope of Calar Alto Observatory in Almer\'ia, Spain. CARMENES has two channels, one in the visible (VIS, 520-960  nm, $R$ =94600) and one in the near-infrared (NIR, 960-1710 nm, $R$ = 80400). We collected two 1800\,s observations per night (separated by around 3 hours) on the four nights between September 28 and October 1, 2023, and four more observations on October 18 and 22, November 24, and December 8, 2023. Two of the observations with the NIR channel  were not usable due to a problem in one of the NIR detectors on the night of September 28. The VIS and NIR spectra taken on the night of October 22 were observed during astronomical twilight, getting both contaminated by the solar spectrum. The raw images were reduced using the \texttt{caracal} pipeline \citep{calablero2016}. The signal-to-noise ratios ranged around 9 and 30 for the VIS and NIR, respectively. We used \texttt{Serval}  \citep{zechmeister2018} to simultaneously determine the $v \sin i_\star$ of the star (using parameter \texttt{-vsiniauto} in \texttt{serval}) and the RVs from the VIS spectra. To determine the  $v \sin i_\star$ we compared the spectrum of our target with template spectra of stars with various spectral types. Using an M4V template (J11477+008/Ross\,128) the single-measurement uncertainties and the RV scatter around the mean were minimised, as compared to using the other templates (M5.5: J00067-075/GJ\,1002; M6.0: J07403-174/GJ\,283\,B; M7.0: J02530+168/Teegarden's star). The $v \sin i_\star$  estimated with that template was $4.2\pm0.4$~km\,s$^{-1}$. Assuming the stellar spin axis aligned to the orbit of planet b, this would result in a rotation period of $1.48\pm0.14\,{\rm d}$. The RV measured from the VIS and NIR spectra have mean uncertainties of 69.0~m\,s$^{-1}$  and 21.9~m\,s$^{-1}$, respectively. Their analysis discards a slope with an amplitude larger than 5 m/s per day over the two months covered by the CARMENES observations.\\

\noindent\textbf{High angular resolution imaging.} SPECULOOS-3 was observed on 2023 October 4 UT using the ‘Alopeke speckle instrument mounted on the Gemini North 8-m telescope \cite{Scott2021,Howell2022}.  ‘Alopeke provides simultaneous speckle imaging in two bands (562\,nm and 832\,nm) with output data products including a reconstructed image with robust contrast limits on companion detections \cite{Howell2016}. Twelve sets of $1000 \times 0.06$-s images were obtained and processed in our standard reduction pipeline \cite{Howell2011}. Extended Data Fig. 2 shows our final contrast curves and the 832-nm reconstructed speckle image. We find that SPECULOOS-3 is a single star to within the angular and contrast levels achieved with no close companions detected brighter than 5--6 magnitudes below that of the target star from $0.1''$ out to $1.2''$. At the distance of SPECULOOS-3 ($d=16.75$\,pc), these angular limits correspond to spatial limits of 1.7 to 20\,AU. It is possible to detect companions down to the diffraction limit of Gemini's 8-m mirror, but at worse contrast. Equal brightness companions are excluded at 20 mas (0.34 AU).

Using the Baraffe models \cite{Baraffe2015}, $\Delta$mag = 5 at 832\,nm, the reddest of both bands, corresponds to a $0.071~ \rm M_\odot$ ($74~\rm M_{jup}$) object, at the maximum age inferred for SPECULOOS-3A (8 Gyr). This mass only leaves sub-stellar objects as plausible companions, unless exactly aligned with the line of sight. If a companion is exactly on the line of sight, it would produce a maximum change in radial-velocity right now.  At a distance of 3.4 AU, a $74~ \rm M_{jup}$ would produce a radial-velocity displacement on SPECULOOS-3A with a semi-amplitude $K_\star = 2.8~ \rm km\,s^{-1}$ over an orbital period $P \sim 15~\rm year$. This radial-velocity displacement is already refuted by the APOGEE observations.\\

\noindent\textbf{Lower limit on the magnitude of a background star.} Thanks to the relatively high proper motion of SPECULOOS-3 (${\approx}0.5''$ per year), we were able to assess presence of a background star at its current position (RA=20h49min27s, Dec. =+$33^{\circ}02'35''$). We used a 2MASS image \citep{Skrutskie2006} obtained in 1998 in $J$ band. We detected no possible additional source at the current position of SPECULOOS-3. The faintest star in the 2MASS image has $J$ magnitude of 16.7. We adopt this value as an absolute lower threshold for the $J$-band magnitude of a background source blended with SPECULOOS-3 in our images obtained with SPECULOOS-North Artemis telescope.  We also compared our current images to DSS digitized archival images taken more than six decades ago. This comparison (Extended Data Fig. 1) did not reveal any background object at the position of the target. \\

\noindent\textbf{Global analysis of the transit photometry.} We performed several global analyses of the transit light curves gathered by SPECULOOS, UKIRT/WFCAM, GTC/HiPERCAM, and MUSCAT-3. These analyses were all done with \texttt{Trafit}, a revised version of the adaptive Markov chain Monte Carlo code presented in  \cite{Gillon2010, Gillon2012, Gillon2014}.  The model assumed for each light curve was
composed of the eclipse model of \cite{MandelAgol2002}, multiplied by a baseline model aiming to represent the other astrophysical and instrumental mechanisms able to produce photometric variations. For each light curve, the baseline model consisted in a polynomial function of external parameters (e.g., time, airmass, x- and y-position, etc.) selected by minimisation of the  Bayesian information criterion (BIC) \cite{Schwarz1978}. 
In our nominal analysis, the stellar mass $M_\ast$, radius $R_\ast$, metallicity [Fe/H], parallax $\pi_\ast$, effective temperature $T_\mathrm{eff}$, and luminosity $L_\ast$ were  kept under the control of the following prior normal  probability distribution functions (PDFs): $N(0.1017, 0.0024^2)\,M_\odot$, $N(0.123, 0.011^2)\,R_\odot$, $N(0.16, 0.07^2)$\,dex, $N(57.701, 0.043^2)$\,mas, $N(2800, 120^2)$\,K, and $N(0.000835, 0.000019^2)\,L_\odot$, respectively. These prior PDFs reflected our a priori knowledge of the stellar properties (see the sections on host star properties above). Given its extremely short orbital period, a circular orbit was assumed for the planet under tidal circularization arguments \cite{Mastumura2008}. A quadratic limb-darkening law \cite{MandelAgol2002} was assumed for the star. For the bandpasses $g'$, $r'$, $i'$, $z'$, and $Ks$,  values and errors for the linear and  quadratic coefficients $u_1$ and $u_2$ were derived from the tables of \cite{Claret2012}, and the corresponding normal distributions were used as prior PDFs in the MCMC (Supplementary Information Fig. 3).  For the non-standard filter $I$+$z$, we adopted the mean value of the filters $i'$ and $z'$ for both coefficients.  For our three analyses, a preliminary MCMC chain of
50,000 steps was first performed to estimate the need to rescale the photometric errors \cite{Gillon2012}. Then a longer MCMC analysis was performed, composed of two chains of 100,000 steps, whose convergence was checked using the statistical test of \cite{Gelman1992}.  In our nominal analysis, we assumed a common transit depth for all bandpasses. The parameters derived from this analysis for the star and its planet are shown in Table 1. Our second analysis assumed different transit depths in each bandpass to check the chromaticity of the transit. The resulting transit depths are shown in Fig. 2. They are all consistent with each other at less than 1 $\sigma$, a decisive element in the confirmation of the planetary nature of SPECULOOS-3\,b.  Finally, we performed a third analysis assuming a non-informative uniform prior on the stellar radius and luminosity to obtain an unbiased measurement of the stellar density from the transit photometry alone. It resulted in a stellar density of $55.1_{-3.7}^{+2.0}\,\rho_\odot$, in excellent agreement with the density of $55 \pm 15\,\rho_\odot$ derived from the a priori knowledge of the star, thus bringing a further validation of the planetary origin of the transit signals.  We also performed individual analyses of the transit light curves to measure the mid-transit timing for each of them (Supplementary Information Fig. 4). A linear regression analysis of these timings/epochs did not reveal any significant deviation from a strictly periodic orbit.  \\

\noindent\textbf{Search for a second planet.} Under the arguments presented by \cite{Gillon2011}, we monitored intensively SPECULOOS-3 with SNO-Artemis and MuSCAT-3 (data described above) in the hope to detect additional transiting planets on longer orbital periods. Our visual inspection of all the gathered light curves did not reveal any convincing additional transit-like structure. After detrending, removal of flares and of transits of SPECULOOS-3\,b, we performed a global analysis of all our ground-based light curves with the \texttt{TLS} algorithm \cite{Hippke2019} that failed to detect any significant power excess indicative of a second transiting planet. We also analyzed the flattened TESS light curve with the \texttt{SHERLOCK} package \citep{pozuelos2020}. We found, at first, a strong signal corresponding to the 0.72\,d candidate, which allowed us to confirm the detectability of this planet in TESS data (see Supplementary Information Fig. 2). Aside from this signal, we found a few other weaker signals that were all refuted by our ground-based observations. We then performed injection and retrieval experiments on this dataset, which allowed us to establish detection limits.  To this end, we used our \texttt{MATRIX} package \citep{pozuelos2020,devora2023}, which generates a sample of synthetic planets by combining a range of orbital periods, planetary radii, and orbital phases that were injected in the  TESS flattened light curve. In particular, we generated 2700 scenarios processed in the search for transits using a process that mimics the \texttt{SHERLOCK} procedure. From the results displayed in Supplementary Information Fig. 5, we conclude that TESS data allows us to detect Earth-size planets with orbital periods shorter than 1\,d. However, the detectability of such small planets rapidly decreases for longer orbital periods, and if they exist and transit, their detection would be very challenging. On the other hand, transiting super-Earth planets with sizes larger than 
1.5\,R$_{\oplus}$ would be easily detectable with recovery rates ranging from 60--100\% for any orbital period up to 10\,d, which allows us to conclude that planets like this do not exist in the system.  \\

\noindent\textbf{Prospect for mass measurements.} Assuming a rocky composition ($5.5\pm0.3\,{\rm g\,cm^{-3}}$) for SPECULOOS-3\,b, the expected mass would be $0.93^{+0.12}_{-0.11}\,{\rm M_\oplus}$. Thanks to its short orbital period (0.72\,d), SPECULOOS-3\,b, is expected to produce a radial-velocity signal with a semi-amplitude $K = 3.1\pm0.4\,\rm m\,s^{-1}$ that should be within reach of stat-of-the-art  high-resolution high-stability spectrographs mounted on 10m-class telescopes. To test this hypothesis, we used (i) the empirical relation or late, slowly rotating M dwarfs, obtained as part of the instrumental commissioning of the MAROON-X instrument \cite{seifahrt20} at the 8.1-m Gemini North telescope (published online at \hyperlink{https://www.gemini.edu/instrumentation/maroon-x/exposure-time-estimation}{https://www.gemini.edu}), and (ii) the empirical linear dependence of the precision with the stellar $v \sin i_\star$ \cite{hatzes10} with as input the CARMENES  measurement, to estimate that a monitoring campaign of 30--45 MAROON-X spectra should result in a 3--4 $\sigma$ detection of the radial velocity signal of SPECULOOS-3\,b and, hence, to a first measurement of its mass.\\

\noindent\textbf{The JWST emission opportunity.} Measuring the emission spectrum of an exoplanet is typically harder than its transmission spectrum. However, emission comes with advantages that overall make it a robust method to study an exoplanet. In particular, emission spectra are not affected by stellar photospheric heterogeneity.  Indeed, in transit, the average stellar spectrum over the transit chord can differ from the rest of the photosphere because of heterogeneities leading to spurious spectral features in the transmission spectrum \citep{2018ApJ...853..122R, Rackham2019, Rackham2023}. As expected, this effect is observed in the case of TRAPPIST-1 b (part of GO 1281 \cite{Lim2023}), which makes inferring an atmosphere for the TRAPPIST-1 planets a challenging task. Besides, UDS are known to be magnetically active with frequent flaring (confirmed from the TESS light curves and ground-based photometric observations). Unfortunately this can be damaging for transmission spectroscopy, as frequent flares visible in the near-IR can affect the shape of the eclipses and the retrieval of the planet parameters (see recommendations from \cite{Lim2023} about the TRAPPIST-1 b transit spectrum). Luckily, the contrast of flares drops with longer wavelengths and their mid-IR counterpart should impact much less the light curves. Emission spectroscopy has other advantages over transmission spectroscopy. First, the interpretation of emission spectra is not dependent on the mass of the planet. Secondly, emission spectra provide the energy budget of the planet, which is essential to understand its atmosphere's chemistry, its dynamics and can be used to constrain the planet's albedo. Finally, in the absence of an atmosphere, emission spectroscopy instead directly accesses the planetary surface where its mineralogy can be studied, something impossible to achieve with transmission spectroscopy. 
For all these reasons, emission spectroscopy is a more reliable method to assess the presence of an atmosphere and study the nature of terrestrial planets around UDS. And, as shown on Fig. 4 (right), SPECULOOS-3\,b is one of the smallest terrestrial planets that is within reach of the JWST in emission spectroscopy with MIRI/LRS.
In that regard, we modeled plausible atmospheric and surface scenarios and compared them to realistic simulations of MIRI/LRS observations of SPECULOOS-3 b in emission.\\

\noindent\textbf{Model emission spectra.} We first modelled the emission spectra of SPECULOOS-3\,b assuming CO$_2$-dominated and H$_2$O-dominated atmospheres, which are among the most plausible atmospheric composition families for hot, rocky planets \cite{Forget:2014}. For this, we performed 3-D global climate model (GCM) calculations using the state-of-the-art Generic Planetary Climate Model (PCM)\cite{Leconte:2013,Turbet2023a}. The model includes self-consistent treatment of radiation, convection and clouds. Parameterizations of these processes are detailed in \cite{Turbet2023a}. In short, radiation is computed using the correlated-$k$ approach, using opacity tables based on HITRAN and HITEMP ; convection is represented through dry and moist adjustment schemes ; H$_2$O and CO$_2$ cloud formation is treated using a prognostic scheme, and assumes a fixed amount of cloud condensation nuclei (CCN). Simulations show a strong temperature inversion in the stratosphere as well as a strong day-night temperature contrast, affecting the emission spectra (see Fig. 5, left). Although the GCM models include cloud formation, no H$_2$O or CO$_2$ clouds form in these simulations, due to the high irradiation the planet receives, which reduces the planet's albedo and thus increases its thermal emission.  We then modelled the theoretical emission spectrum of an airless planet with a null-albedo and absence of heat redistribution by considering the planetary flux to be a sum of black bodies calculated for a grid of $T_{\theta,\phi}$ where $\theta$ and $\phi$ are the longitude and the latitude respectively. The temperature of the sub-stellar point (at zenith) is chosen to be the maximal dayside temperature defined as $T_{\rm day,~max}=T_\star \times \sqrt{\frac{R_\star}{a}}\times(\frac{2}{3})^{1/4}$. The temperature then decreases with increasing latitude and longitude.  Besides this simple blackbody model, we also modeled emission spectra for several geologically plausible planetary surface types, including ultramafic crust formed from solidification of a magma ocean or high-temperature lava flows, basaltic crust that forms from volcanic eruptions like those on present-day Earth, and granitoid crust formed from crustal re-processing. The data for different surfaces are based on single scattering albedos derived by \cite{2012ApJ...752....7H}. An ultramafic crust and a granitoid crust will be mainly composed of silicates that have intense spectral features between 8 and 12 microns due to Si--O stretching \cite{Salisbury1993}, and these features are detectable with MIRI/LRS observations (see PandExo simulations below and Fig. 5, right).  In addition to searching for spectral features, emission spectroscopy can also constrain the planet's dayside albedo and thus its geological history. In the Solar System, airless bare rocks like the Moon and Mercury are much darker than fresh geologic surfaces due to the effect of space weathering \cite{2001JGR...10610039H}. Should SPECULOOS-3\,b be an airless rocky planet, then we would similarly expect its surface to become progressively darker. We performed simulations of this effect by considering the albedo effect of graphite particles on a fresh ultramafic surface, using the same approach as in \cite{Lyu2023}. We find that even moderate space weathering would substantially lower the dayside albedo, thereby increasing the planet's brightness temperature (and thus eclipse depth) at MIRI/LRS wavelengths (see Extended Data Fig. 4). For reference, the space weathering timescale on TRAPPIST-1\,c was estimated to be 10$^2$--10$^3$ years \cite{2023Natur.620..746Z}; due to their similar planetary and stellar parameters, the space weathering timescale for SPECULOOS-3\,b should be of a similar order of magnitude. Any detection of a high-albedo dayside would thus require that the planet's surface is geologically very young, indicating that the planet has a very active volcanism or tectonic overturn.\\

\noindent\textbf{PandExo simulations.} We used the \texttt{PandExo} \cite{batalha_pandexo_2017} simulation tool to estimate the noise of a single secondary eclipse observation of SPECULOOS-3\,b with MIRI/LRS in slitless spectroscopy mode. For the stellar spectrum, we used the best-fit PHOENIX model from our SED fitting (see above), normalized to a $J$-mag = 11.501. We assumed 180 groups per integration in order to optimize the observing efficiency ($\sim$99\%), while keeping the counts level below 90\% saturation according to JWST ETC. This yields a total time per integration (including readout and reset) of 28.63 sec. We set the total observing time to 2.9 hours. This duration takes into account a margin of 1 hour for the start of the observations, 30 minutes for the stabilisation of the detector, 28 minutes of baseline before the start of the eclipse, 28 minutes during the eclipse, and 28 minutes of baseline after the eclipse. Finally, we assumed a systematic noise floor of 30 ppm based on the performances reported from previous observations (e.g., \cite{2023PASP..135c8002B}). We then tested how many secondary eclipse observations would be necessary to distinguish different models from one another, following an approach similar to that of \cite{2022AJ....164..258W}. For each pair of models, we conducted a chi-square test and obtained a $p$-value representing the probability that the simulated data from the model to be tested are consistent with the second model. For each simulated dataset, we used the PandExo output that included a random Gaussian noise component (\texttt{`spectrum\_w\_rand'} key in the output file). We started with one eclipse and increased the number of eclipses until the two models were distinguishable by 4 $\sigma$, corresponding to a $p$-value of less than 0.000063. We conducted these tests using a variety of wavelength bins, either of constant resolution $R$ ($\lambda$/$\Delta \lambda$) or constant widths ($\Delta \lambda$). We found that the models are maximally distinguishable for a resolution $R$=3 in the first case, or a bin width = 2$\mu$m in the second case.  Extended Data Table 2 shows the number of secondary eclipse observations needed to distinguish the CO$_2$ and H$_2$O atmospheric models from two common airless planet models (blackbody and basaltic) at 4 $\sigma$. At most, 7 eclipses are needed to assess the presence of the most plausible atmosphere for SPECULOOS-3\,b with 4 $\sigma$ confidence. Extended Data Table 2 also shows that observing two additional occultations, i.e., 9 occultations in total, should allow to distinguish a CO$_2$-dominated atmosphere from an H$_2$O-dominated one. Should there be no evidence for an atmosphere, the MIRI/LRS emission spectrum could be used to assess different surface compositions. Extended Data Table 3 reports the number of secondary eclipse observations needed to distinguish different surface models at 4 $\sigma$. About 50\% of the considered surface pairs should be distinguishable at $\geq$ 4 $\sigma$ with 10 eclipse observations.

\newpage
\backmatter

\bmhead{Data availability} All the data (Kast, SpeX, and CARMENES spectra; SPECULOOS, Saint-EX, T150, MuSCAT3, GTC/HiPERCAM, UKIRT/WFCAM, TESS light curves) used in this work are publicly available on the following Zenodo online repository: \url{https://doi.org/10.5281/zenodo.10821723}

\bmhead{Code availability} The code \texttt{PROSE} used to reduce the SPECULOOS, TRAPPIST, and MuSCAT3 data is available at \url{https://github.com/lgrcia/prose}.  
The code \texttt{TRAFIT} used to analyse the light curves is a Fortran 2003 code that can be obtained from the first author on reasonable request.  
The {\tt HiPERCAM} pipeline is available at \url{https://cygnus.astro.warwick.ac.uk/phsaap/hipercam/docs/html/}.
The \texttt{SHERLOCK} package used to search for planets in the TESS data is publicly available at  
\url{https://github.com/franpoz/SHERLOCK}. 
The detection limits in the TESS data were computed using the \texttt{MATRIX} package, publicly available at \url{https://github.com/PlanetHunters/tkmatrix}. 
The code used to create Extended Data Fig. 1 is available at \url{https://github.com/jpdeleon/epoch}.
The Generic Planetary Climate Model code (and documentation on how to use the model) used in this work can be downloaded from the SVN repository at \url{https://svn.lmd.jussieu.fr/Planeto/trunk/LMDZ.GENERIC/}. More information and documentation are available at \url{http://www-planets.lmd.jussieu.fr}.
The code \texttt{kastredux} used to reduce the Kast optical spectrum is available at \url{https://github.com/aburgasser/kastredux}. 

\bmhead{Acknowledgments}
The ULiege's contribution to SPECULOOS has received funding from the European Research Council under the European Union's Seventh Framework Programme (FP/2007-2013) (grant Agreement n$^\circ$ 336480/SPECULOOS), from the Balzan Prize and Francqui Foundations, from the Belgian Scientific Research Foundation (F.R.S.-FNRS; grant n$^\circ$ T.0109.20), from the University of Liege, and from the ARC grant for Concerted Research Actions financed by the Wallonia-Brussels Federation. 
M.G. is F.R.S-FNRS Research Director. His contribution to this work was done in the framework of the PORTAL project funded by the Federal Public Planning Service Science Policy  (BELSPO) within its  BRAIN-be: Belgian Research Action through Interdisciplinary Networks program. E.J. is F.R.S-FNRS Senior Research Associate. VVG is F.R.S-FNRS Research Associate. 
The postdoctoral fellowship of KB is funded by F.R.S.-FNRS grant T.0109.20 and by the Francqui Foundation. This publication benefits from the support of the French Community of Belgium in the context of the FRIA Doctoral Grant awarded to M.T.
This work is supported by a grant from the Simons Foundation (PI Queloz, grant number 327127). J.d.W. and MIT gratefully acknowledge financial support from the Heising-Simons Foundation, Dr. and Mrs. Colin Masson and Dr. Peter A. Gilman for Artemis, the first telescope of the SPECULOOS network situated in Tenerife, Spain. B.-O. D. acknowledges support from the Centre for Space and Habitability of the University of Bern and the Swiss State Secretariat for Education, Research and Innovation (SERI) under contract number MB22.00046. Part of this work received support from the National Centre for Competence in Research PlanetS, supported by the Swiss National Science Foundation (SNSF). The Birmingham contribution research is in part funded by the European Union's Horizon 2020 research and innovation programme (grant's agreement n$^{\circ}$ 803193/BEBOP), from the MERAC foundation, and from the Science and Technology Facilities Council (STFC; grant n$^\circ$ ST/S00193X/1, and ST/W000385/1).
The authors thank the Belgian Federal Science Policy Office (BELSPO) for the provision of financial support in the framework of the PRODEX Programme of the European Space Agency (ESA) under contract number 4000142531.
B.V.R. thanks the Heising-Simons Foundation for Support.
This material is based upon work supported by the National Aeronautics and Space Administration under Agreement No.\ 80NSSC21K0593 for the program ``Alien Earths''.
The results reported herein benefited from collaborations and/or information exchange within NASA’s Nexus for Exoplanet System Science (NExSS) research coordination network sponsored by NASA’s Science Mission Directorate.
M.Tu. acknowledges support from the Tremplin 2022 program of the Faculty of Science and Engineering of Sorbonne University. M.Tu. thanks the Generic PCM team for the teamwork development and improvement of the model. M.Tu. acknowledges support from the High-Performance Computing (HPC) resources of Centre Informatique National de l'Enseignement Supérieur (CINES) under the allocations No. A0100110391, A0120110391 and A0140110391 made by Grand Équipement National de Calcul Intensif (GENCI).
M.R.S acknowledges support from the
European Space Agency as an ESA Research Fellow.
E.A.M.V. acknowledges support from the Centre for Space and Habitability
(CSH). This work has been carried out within the framework of the National
Centre of Competence in Research PlanetS supported by the Swiss National
Science Foundation under grants 51NF40\_182901 and 51NF40\_205606.
This work is based upon observations carried out at the Observatorio Astron\'omico Nacional on the Sierra de San Pedro M\'artir (OAN-SPM), Baja California, M\'exico.
SAINT-EX observations and team were supported by the Swiss National Science Foundation (PP00P2-163967 and PP00P2-190080),
 the Centre for Space and Habitability (CSH) of the University of Bern,  the National Centre for Competence in Research PlanetS, supported by the Swiss National Science Foundation (SNSF). 
Y.G.M.C acknowledges support from UNAM PAPIIT-IG101224.
Based on observations made with the GTC telescope, in the Spanish Observatorio del Roque de los Muchachos of the Instituto de Astrofísica de Canarias, under Director’s Discretionary Time.
Some of the observations in this paper made use of the High-Resolution Imaging instrument ‘Alopeke and were obtained under Gemini LLP Proposal Number: GN-2023B-DD-101. ‘Alopeke was funded by the NASA Exoplanet Exploration Program and built at the NASA Ames Research Center by Steve B. Howell, Nic Scott, Elliott P. Horch, and Emmett Quigley. Alopeke was mounted on the Gemini North telescope of the international Gemini Observatory, a program of NSF’s OIR Lab, which is managed by the Association of Universities for Research in Astronomy (AURA) under a cooperative agreement with the National Science Foundation. on behalf of the Gemini partnership: the National Science Foundation (United States), National Research Council (Canada), Agencia Nacional de Investigación y Desarrollo (Chile), Ministerio de Ciencia, Tecnología e Innovación (Argentina), Ministério da Ciência, Tecnologia, Inovações e Comunicações (Brazil), and Korea Astronomy and Space Science Institute (Republic of Korea).
F.J.P., P.J.A., A.S., R.V., and J.A. acknowledge financial support from the grant CEX2021-001131-S funded by MCIN/AEI/ 10.13039/501100011033.
This work benefits from observations made at the Sierra Nevada Observatory, operated by the Instituto de Astrof\'isica de Andaluc\'ia (IAA-CSIC).
D.K. and X.L. acknowledge financial support from NSFC grant 42250410318.
F.S. acknowledges support from CNES, Programme National de Planétologie (PNP)and of the Investments for the Future programme IdEx, Université de Bordeaux/RRI ORIGINS. 
This research was carried out in part at the Jet Propulsion Laboratory, California Institute of Technology, under a contract with the National Aeronautics and Space Administration (80NM0018D0004).
This work is partly supported by MEXT/JSPS KAKENHI Grant Numbers
JP15H02063, JP18H05439, JP18H05442, JP21K13955, JP21K20376,
JP22000005, and JST CREST Grant Number JPMJCR1761.
This paper is based on observations made with the MuSCAT3 instrument,
developed by the Astrobiology Center and under financial supports by
JSPS KAKENHI (JP18H05439) and JST PRESTO (JPMJPR1775), at Faulkes
Telescope North on Maui, HI, operated by the Las Cumbres Observatory.
Observations made with the Wide-Field Camera (WFCam) on the UKIRT telescope were granted through Director’s Discretionary Time. UKIRT is owned by the University of Hawaii (UH) and operated by the UH Institute for Astronomy. 
This research has made use of the NASA Exoplanet Archive, which is operated by the California Institute of Technology, under contract with the National Aeronautics and Space Administration under the Exoplanet Exploration Program.
The Digitized Sky Surveys were produced at the Space Telescope Science Institute under U.S. Government grant NAG W-2166. The images of these surveys are based on photographic data obtained using the Oschin Schmidt Telescope on Palomar Mountain and the UK Schmidt Telescope. The plates were processed into the present compressed digital form with the permission of these institutions.
The National Geographic Society - Palomar Observatory Sky Atlas (POSS-I) was made by the California Institute of Technology with grants from the National Geographic Society.
The Second Palomar Observatory Sky Survey (POSS-II) was made by the California Institute of Technology with funds from the National Science Foundation, the National Geographic Society, the Sloan Foundation, the Samuel Oschin Foundation, and the Eastman Kodak Corporation.
The Oschin Schmidt Telescope is operated by the California Institute of Technology and Palomar Observatory.
The UK Schmidt Telescope was operated by the Royal Observatory Edinburgh, with funding from the UK Science and Engineering Research Council (later the UK Particle Physics and Astronomy Research Council), until 1988 June, and thereafter by the Anglo-Australian Observatory. The blue plates of the southern Sky Atlas and its Equatorial Extension (together known as the SERC-J), as well as the Equatorial Red (ER), and the Second Epoch [red] Survey (SES) were all taken with the UK Schmidt.
All data are subject to the copyright given in the copyright summary. Copyright information specific to individual plates is provided in the downloaded FITS headers.
Supplemental funding for sky-survey work at the STScI is provided by the European Southern Observatory. 

\section{Author Contributions Statement}
M.G.\ initiated the SPECULOOS project, performed the analyses of the photometry described in this paper, and wrote a large part of the paper.
M.G., B.-O.D., J.d.W., D.Q., and A.H.M.T.\ lead the SPECULOOS project and manage its funding, its organisation, and its operations.
P.P.P.\ developed and maintains the SPECULOOS database and its web interface, a key element in the discovery of the planet. 
B.V.R.\ and C.A.T.\ acquired, reduced, and analyzed the SpeX spectra, and B.V.R.\ analyzed the TESS data with J.d.W.
E.D.\ manages the scheduling of the SPECULOOS observations.
A.Y.B.\ manages operations of the SPECULOOS North facility.
S.Z.-F. manages the SPECULOOS South Observatory facilities.
M.H.\ manages the SPECULOOS data analysis pipeline. This role was before in the hands of C.M. (who developed the pipeline).\ 
M.G., A.H.M.T., B.D., S.D., G.D., Y.T.D., M.R.S., T.B., M.H., S.J.T., C.J.M., E.D., K.B., P.P.P., A.Y.B., L.D., M.Ti, F.J.P., S.Z.-F., E.J., L.G., C.M., P.N. operated the SPECULOOS telescopes.\
E.D., G.D., L.D., D.S., M.Ti., F.P., S.Z.-F.\ have examined the SPECULOOS light curves on a daily basis to search for any structure that could be related to the transit of an exoplanet, and identified the first transits of the planet. 
L.G. contributed to the data management and maintenance of the SPECULOOS different observatories.
K.G.S.\ performed the spectral energy distribution analysis.
A.J.B.\ and R.G.\ acquired, reduced, and analyzed the Kast optical spectrum.
C.A.\ led the kinematic and metallicity age analysis.
S.B.H.\ obtained the Gemini high-resolution speckle observations, reduced the data, and provided their analysis. In addition, provided comments on the manuscript.
Z.B.\ managed the Oukaimeden Observatory hosting TRAPPIST-North. Mo. G.\ scheduled and performed the TRAPPIST-North observations.
K.B.\ provided SPECULOOS, Saint-EX, TRAPPIST-North and MuSCAT3 data reduction.
N.N.\ obtained MuSCAT3 Director's Discretionary Time and performed all MuSCAT3 observations.
J.P.d.L.\ reduced data from MuSCAT2 and a part of MuSCAT3.
N.N., J.P.d.L., A.F., I.F., Y.H., K.I., K.I., M.I., K.K., T.K., Y.K., J.H.L., M.M., M.T., Y.T., N.W.\ provided MuSCAT3 GTO for this project.
N.N. and E.P.\ provided MuSCAT2 observations for this project.
F.M.\ performed MuSCAT2 observation.
C.A.C. provided the `Alopeke data reduction. R.A. provided GTC/HiPERCAM data reduction.
F.J.P.\ searched for extra planets in the TESS data and established detection limits. Scheduled, reduced, and analyzed the T150 photometric data, and provided the CARMENES data.
P.J.A., J.A., R.V.\ PIs of proposal, managing the schedule of observations and data reduction and analysis of CARMENES data.
D.S.\ assessed the potential to measure the planets mass using high-resolution spectroscopy.
A.S.\ operated the T150 telescope.
L.D. and E.D.\ assessed the potential of the planet for emission spectroscopy with JWST and performed the corresponding PandExo simulations.
R.H., D.D.B.K., and X.L.\ provided model emission spectra for various surface compositions. 
M.Tu. and F.S.\ performed 3-D numerical climate model simulations of SPECULOOS-3b for two plausible atmospheres, and provided associated emission spectra.
E.B. provided guidance and comments on the manuscript.
S.J.T. helped in the design and commissioning of the SPECULOOS South Observatory and is a member of the operations team.
F.S. worked on synthetic observations with JWST.
J. McC. implemented a custom autoguiding routine for each SPECULOOS node based on his open source Donuts science frame autoguiding algorithm (\url{https://github.com/jmccormac01/Donuts, https://pypi.org/project/donuts/}), with the goal of minimising star drift and systematic noise in the observations.
M.J.H. mainlines and develops the pipeline that automatically processes all SPECULOOS data at the University of Cambridge and provided an initial fit to the SPECULOOS and TESS transit light curves.
V.V.G. ran stellar evolution modeling to derive host star properties.
B.O.D., U.S., Y.G.M.C, E.A.M.V.,  I.P.F., L.S., N.S., and F.Z.L. have operated the SAINT-EX telescope, including scheduling observations, nightly operations, data processing, and maintenance of the facility. 
K. H. contributed to the funding of SAINT-EX.

\section{Competing Interests Statement}
The authors declare no competing interests.

\section{Figures and tables}

\begin{figure}[H]
    \centering
   \includegraphics[width=\textwidth]{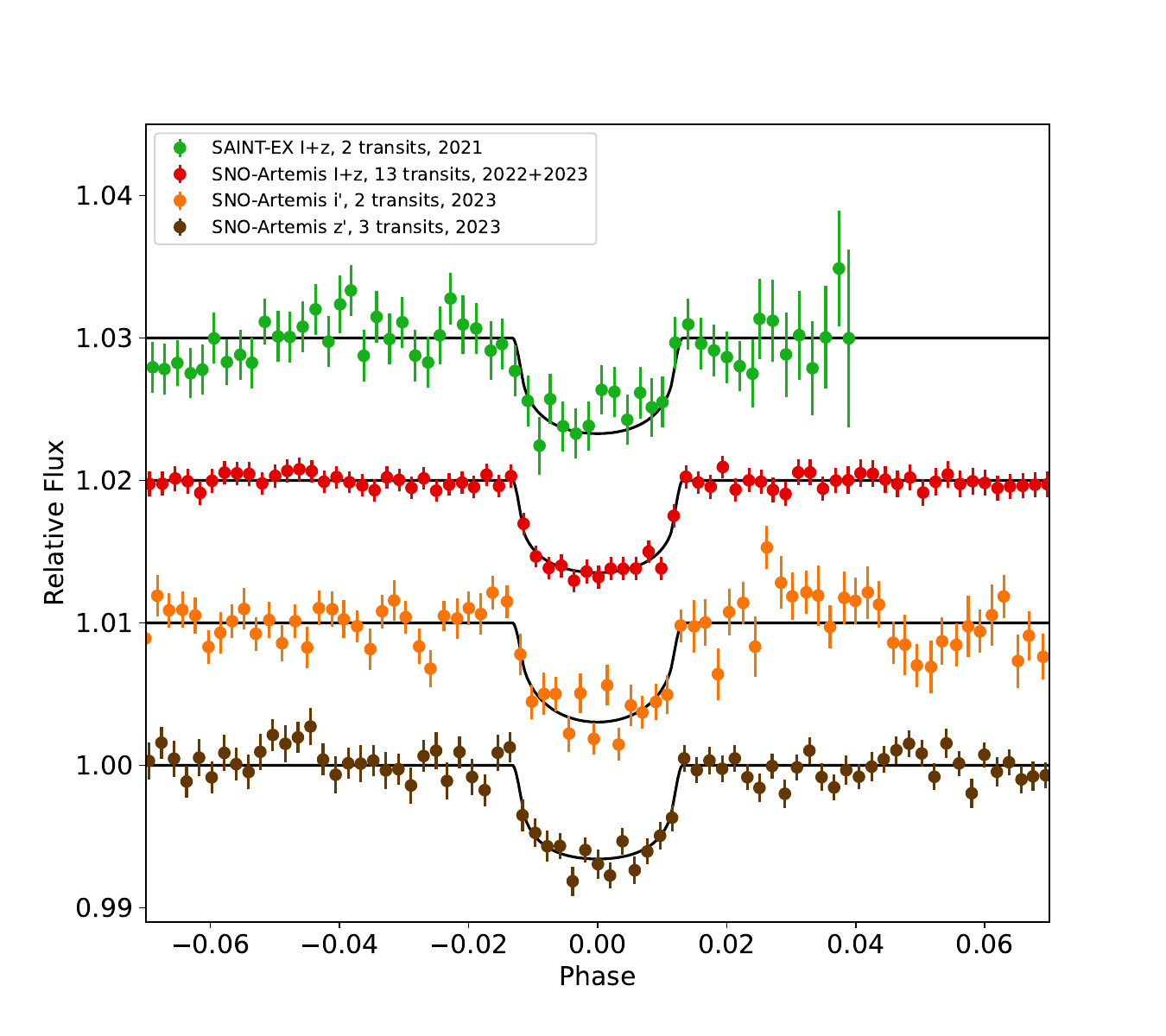}
   \captionsetup{labelformat=empty}
    \caption{\textbf{Fig. 1 $\vert$ Discovery transit photometry of SPECULOOS-3\,b. }Transit photometry obtained by SAINT-EX and SNO-Artemis between 2021 and 2023 (phase-folded, detrended and binned per 2 minutes). The light curves are shifted along the $y$-axis for clarity. The SAINT-EX, SNO $I+z$, $i'$, and $z'$ measurements are, on average, the means of 2.8, 21.6, 6.0, and 2.5 photometric data points. The error bars are the mean errors of the points within the bin divided by the square root of the number of points. The black lines show the best-fit transit models.}
    \label{fig:fig1}
\end{figure}

\begin{figure}[H]
    \centering
   \includegraphics[width=\textwidth]{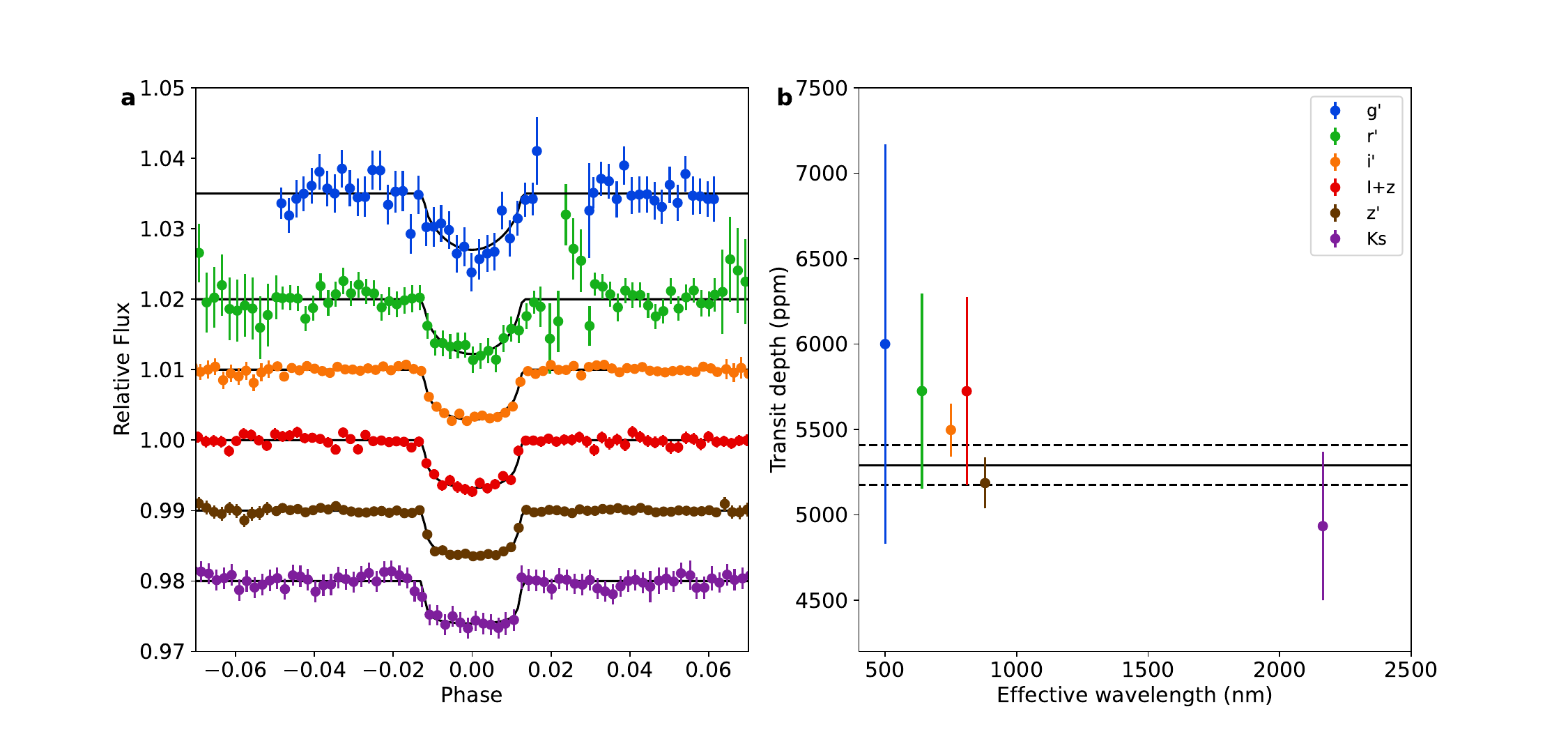}
   \captionsetup{labelformat=empty}
    \caption{\textbf{Fig. 2 $\vert$ Optical and infrared transit photometry of SPECULOOS-3\,b.} \textbf{a)} Phase-folded transit photometry of SPECULOOS-3\,b in (from top to bottom) the $g'$, $r'$, $i'$, $I+z$, $z'$, and $Ks$ filters. The photometry is binned per 2 minutes and the light curves are shifted along the $y$-axis for clarity. The $g'$, $r'$, $i'$, and $z'$ light curves are dominated by the extremely precise GTC/HiPERCAM light curves of only one transit, while the $Ks$ light curve is the stack of two transits observed by UKIRT/WFCAM in the $Ks$ filter. The $g'$, $r'$, $i'$, $I+z$, $z'$, and $Ks$ measurements are, on average, the means of 7.6, 5.3, 17.4, 16.2, 26.9, and 17.7 photometric data points. The error bars are the mean errors of the points within the bin divided by the square root of the number of points. The black lines show the best-fit transit models. \textbf{b)} Measured transit depths in the different filters (dots with error bars). The measurements and the errors are, respectively, the means and the standard deviations of the posterior probability distributions derived from the global MCMC analysis. The measurements are compared to the transit depth measured in the global MCMC analysis assuming a common transit depth (dashed lines = 1 $\sigma$ error bar). All measurements agree at better than 1 $\sigma$ with the common transit depth measurement.}
    \label{fig:fig2}
\end{figure}

\begin{figure}[H]
    \centering
   \includegraphics[width=\textwidth]{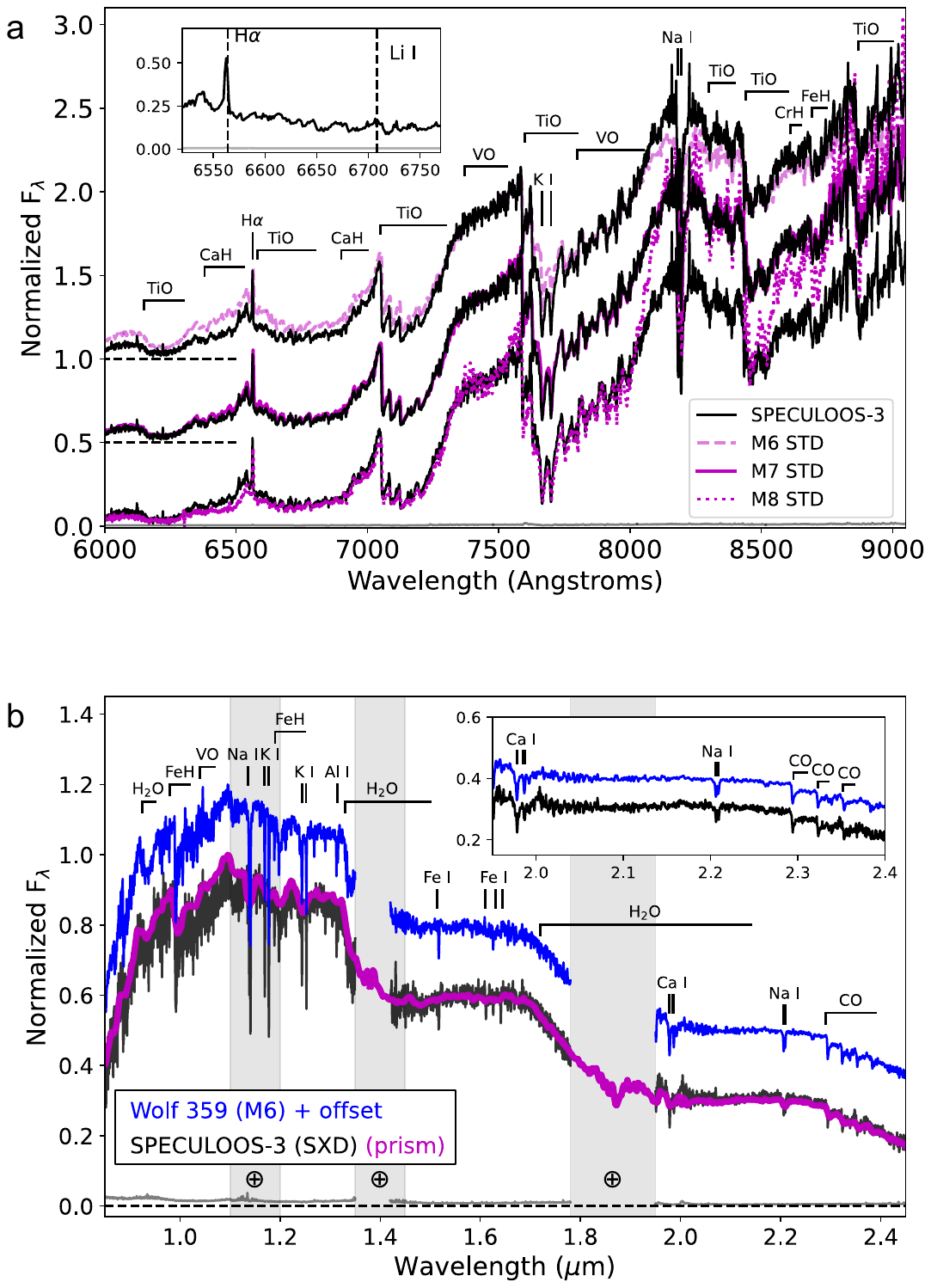}
   \captionsetup{labelformat=empty}
    \caption{\textbf{Fig. 3 $\vert$ Optical and infrared spectroscopy of SPECULOOS-3.} \textbf{a)}  normalized Kast red-optical spectrum of SPECULOOS-3 (black lines) compared to M6, M7 and M8 spectral templates from \cite{2017ApJS..230...16K} (magenta lines in different line styles), with a M7 spectral type providing the best-fit (see Methods). Individual comparisons are offset vertical in steps of 0.5, and the spectral flux uncertainties are indicated by the grey line along the bottom. Prominent line and molecular features in the optical spectra of late-M dwarfs are labeled, and the inset box shows a close-up of the region encompassing H$\alpha$ 6563~{\AA} emission (detected) and Li~I 6708~{\AA} absorption (not detected). \textbf{b)} normalized SpeX near-infrared spectrum of SPECULOOS-3 (SXD data in black, prism data in magenta) compared with a medium-resolution spectrum of the M6 dwarf Wolf 359 (blue line, offset by 0.2 flux units). Prominent spectral features of M dwarfs are noted, and regions of high telluric absorption ($\oplus$) are shaded grey. The grey line at the bottom of the plot illustrates the uncertainties for the target spectrum. The inset box highlights the $K$-band region and its metallicity-sensitive Ca~I, Na~I and CO features, which are nearly identical between SPECULOOS-3 and Wolf~359.}
    \label{fig:fig3}
\end{figure}

\begin{figure}[H]
    \centering
   \includegraphics[width=\textwidth]{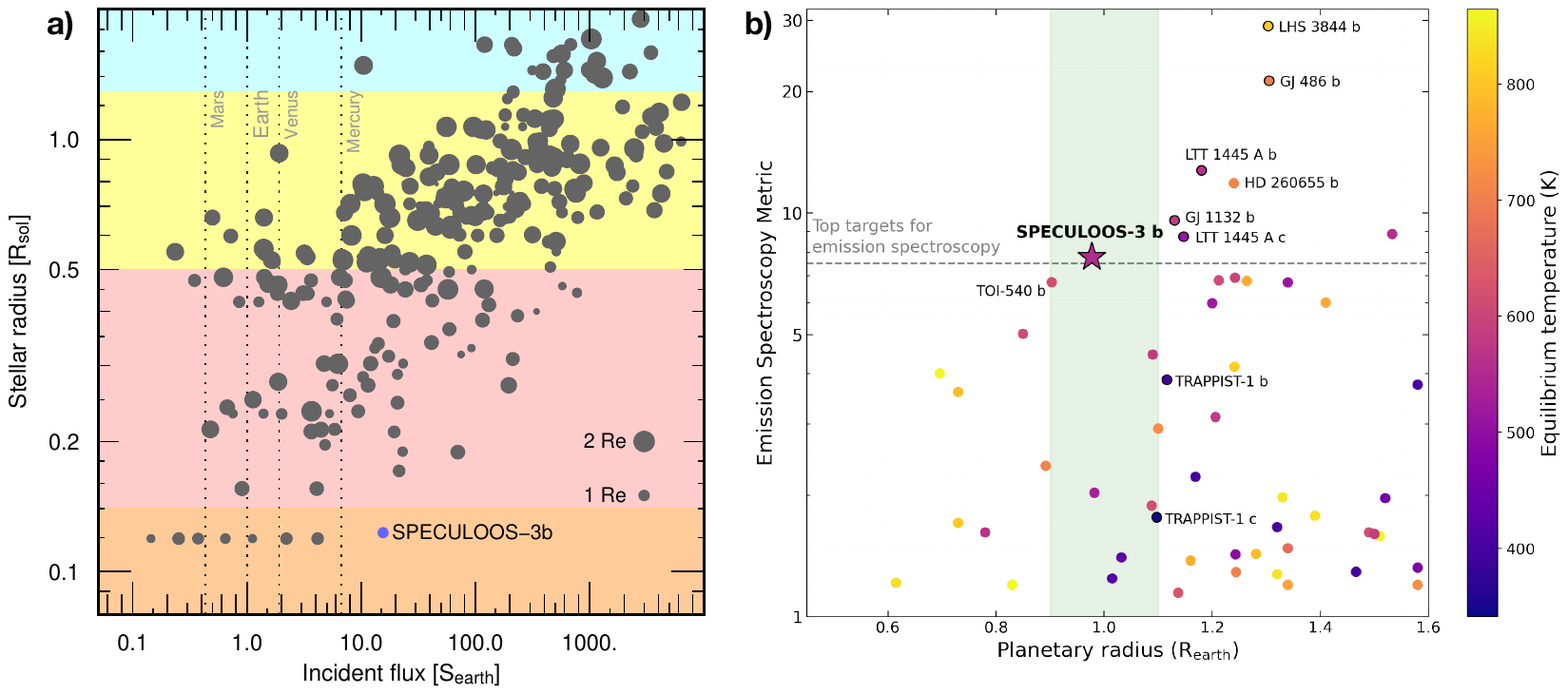}
   \captionsetup{labelformat=empty}
    \caption{\textbf{Fig. 4 $\vert$ Comparison of SPECULOOS-3\,b with other rocky exoplanets.} \textbf{a)} Sizes of host stars and incident stellar flux of known sub-Neptune-sized exoplanets. The size of the symbols scales linearly with the radius of the planet. The background is colour-coded according to stellar size (in units of the Sun’s size), with the ultra-cool dwarf regime shown in orange. The positions of the Solar System terrestrial planets are shown for reference. One can see on this Figure that SPECULOOS-3\,b extends the unique planet sample of TRAPPIST-1 to a larger stellar flux. \textbf{b)} SPECULOOS-3\,b in the context of other known transiting terrestrial exoplanets ($R_{\rm{p}}<1.6 R_\oplus$) that are cool enough ($T_{\rm{eq}}<$ 880 K \cite{2019ApJ...886..141M}) to have a dayside made of solid rock (in contrast to hotter magma worlds with molten surfaces). The planets are shown as a function of their radius and their Emission Spectroscopy Metric \cite{kempton_framework_2018}. They are color-coded as a function of their equilibrium temperature. The shaded green area highlights planetary radii most similar to Earth's (0.9--1.1 $R_\oplus$). The dashed horizontal line represents the threshold of ESM=7.5 recommended by ref.\,\cite{kempton_framework_2018} to identify the top targets for emission spectroscopy with JWST. SPECULOOS-3\,b is one of the smallest planets amenable to emission spectroscopy with MIRI/LRS. Data from \url{https://exoplanetarchive.ipac.caltech.edu} (16th of October 2023).}
    \label{fig:fig4}
\end{figure}

\begin{figure}[H]
    \centering
   \includegraphics[width=\textwidth]{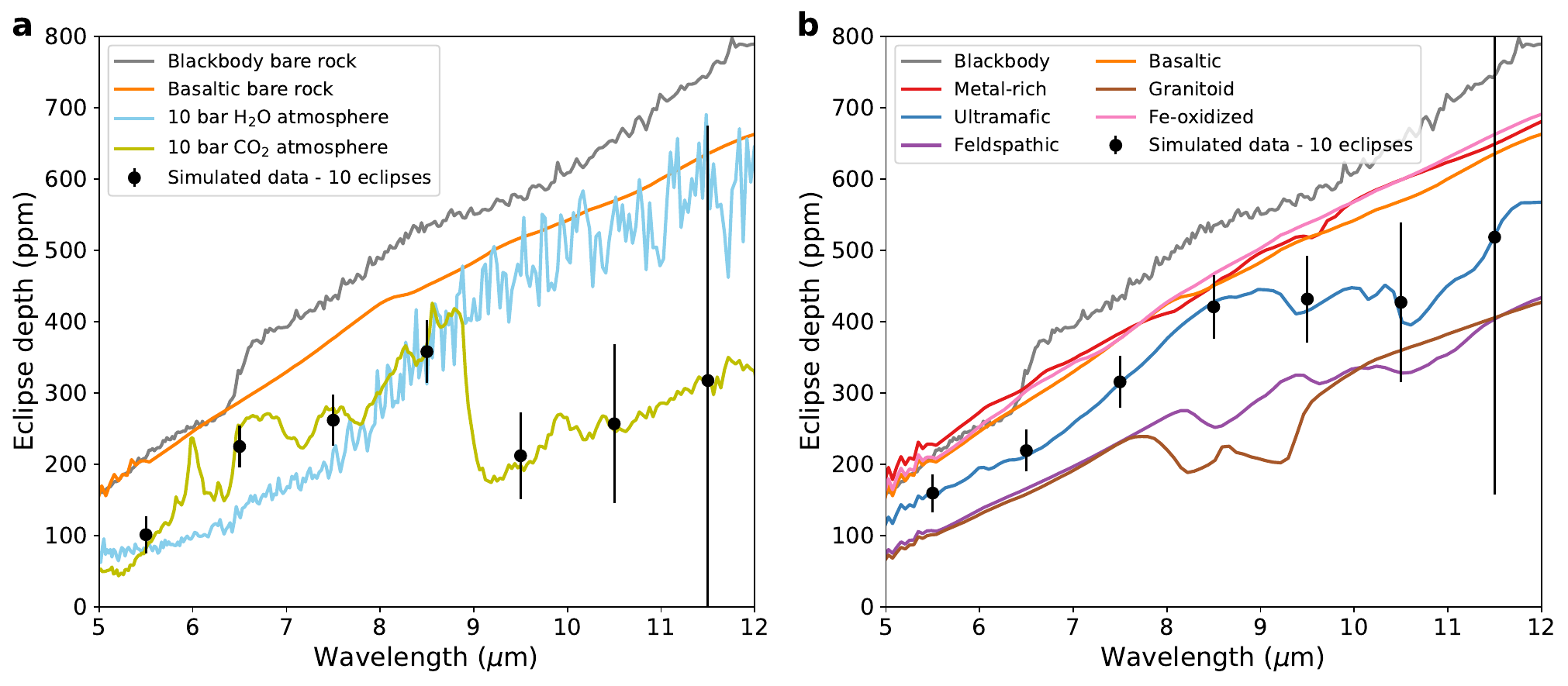}
   \captionsetup{labelformat=empty}
    \caption{\textbf{Fig. 5 $\vert$ Simulated JWST MIRI/LRS emission spectra of SPECULOOS-3b.} \textbf{a)} Model emission spectra of SPECULOOS-3\,b for two plausible atmospheres as well as two bare-surface scenarios for comparison. Simulated MIRI/LRS data are shown for the 10\,bar CO$_2$ atmosphere model. These simulated data were obtained with \texttt{PandExo} \cite{batalha_pandexo_2017} assuming 10 occultation observations with MIRI/LRS. They are shown here binned with a constant binwidth of 1\,$\mu$m (average number of points per bin = 35) and without any Gaussian scatter around the model, for visual clarity.  The error bars are the 1-$\sigma$ errors computed by \texttt{PandExo}. \textbf{b)} Model emission spectra for a range of surface compositions. The simulated MIRI/LRS data are shown here for the ultramafic surface model as an example. Here too, the simulated measurements are binned with a constant binwidth of 1\,$\mu$m and without any Gaussian scatter around the model, for visual clarity, and the error bars are the 1-$\sigma$ errors computed by \texttt{PandExo}.}
    \label{fig:fig5}
\end{figure}

\begin{figure}[H]
    \centering
   \includegraphics[width=\textwidth]{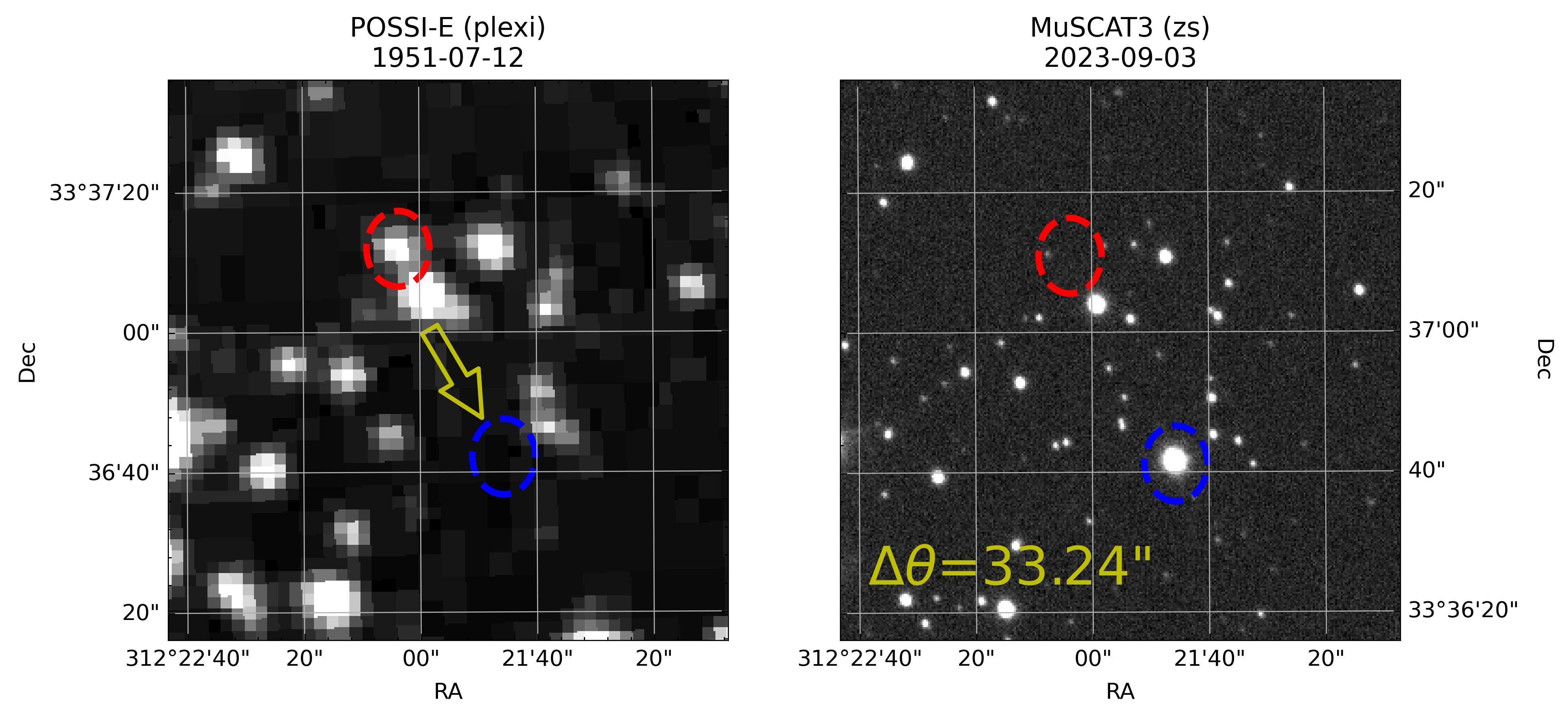}
   \captionsetup{labelformat=empty}
    \label{fig:ed_fig1}
\end{figure}

\begin{figure}[H]
    \centering
   \includegraphics[width=\textwidth]{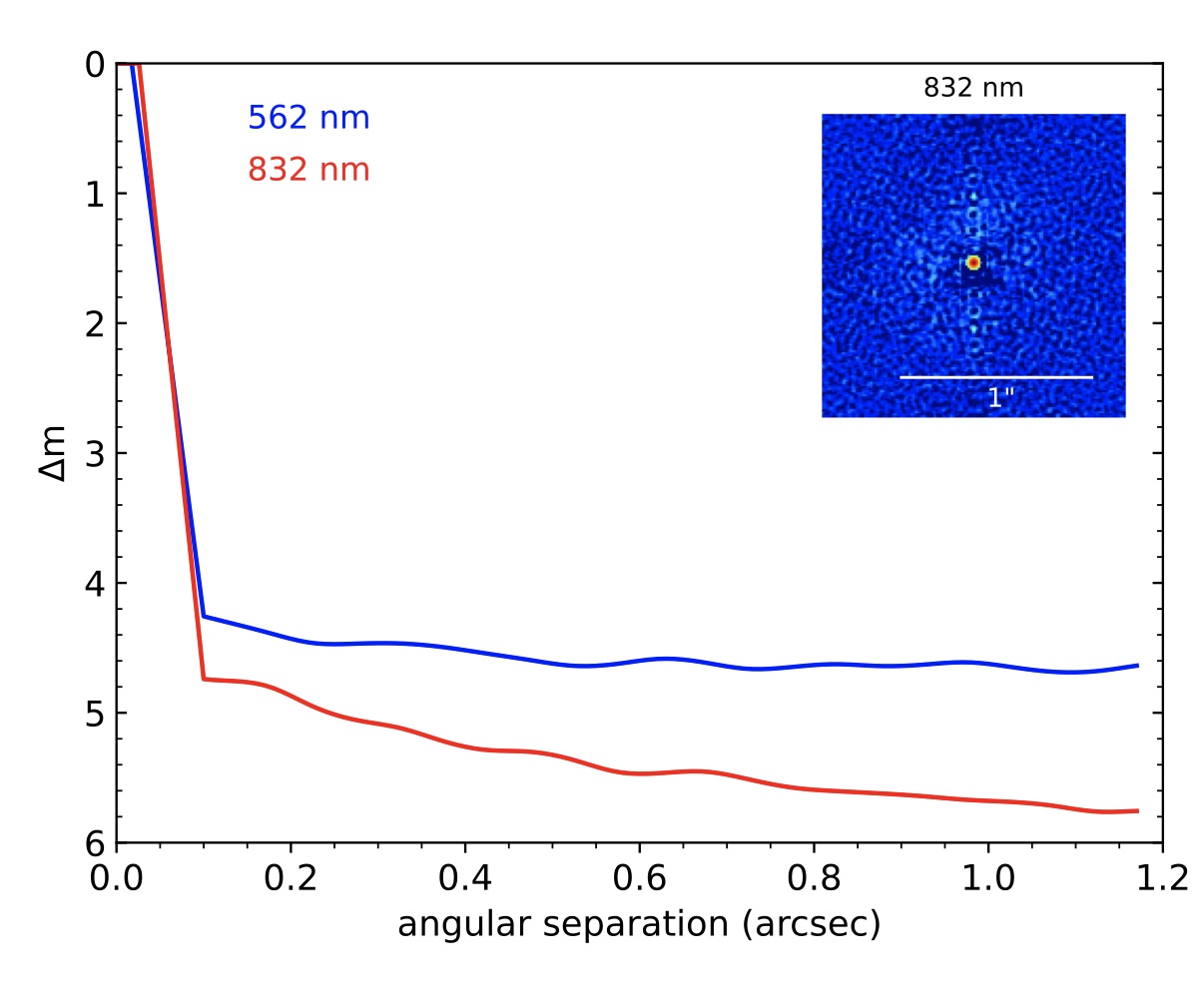}
   \captionsetup{labelformat=empty}
    \caption{\textbf{Extended Data Fig. 2 $\vert$ Speckle imaging of SPECULOOS-3.} Result from speckle imaging with the 'Alopeke instrument mounted on the 8-m Gemini-North telescope, on Maunea Kea, Hawai'i. The inset on the top right shows the final image produced by our analysis, which is summarised by the two curves of the main figure. These curves show the sensitivity in two bands (blue = 562\,nm and red = 832\,nm). The observations reveal there are no companions with a brightness greater than 5 to 6 magnitudes at distances above $0.1''$ from SPECULOOS-3A, which corresponds to a physical distance of approximately 1.7\,AU.}
    \label{fig:ed_fig2}
\end{figure}

\begin{figure}[H]
    \centering
   \includegraphics[width=\textwidth]{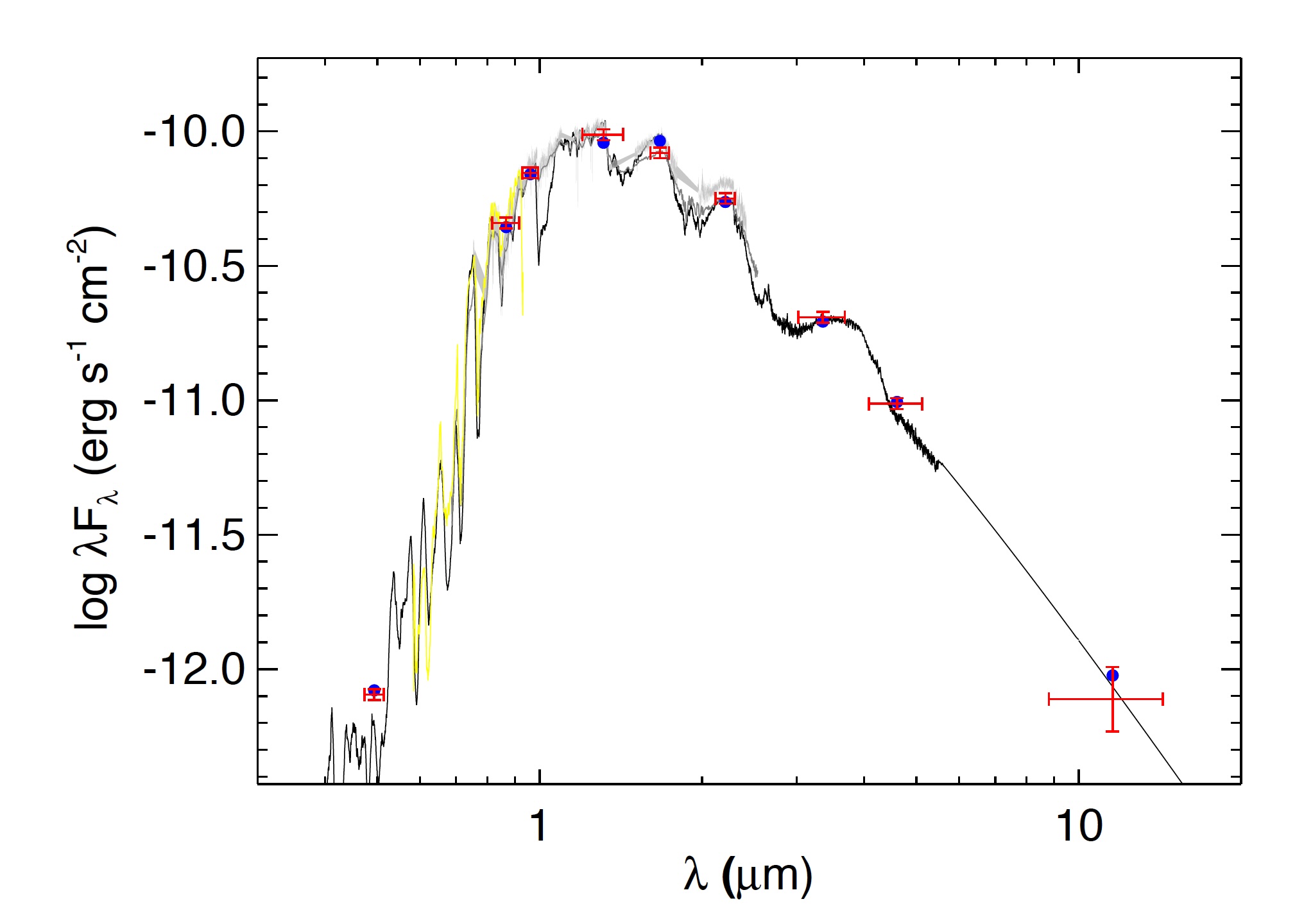}
   \captionsetup{labelformat=empty}
    \caption{\textbf{Extended Data Fig. 3 $\vert$  Spectral energy distribution of SPECULOOS-3.} Red symbols represent the observed photometric measurements, where the horizontal bars represent the effective width of the bandpass, and the vertical bars the 1-$sigma$ error bars on the measurements. Blue symbols are the model fluxes from the best-fit PHOENIX atmosphere model (black). Overlaid on the model are the absolute flux-calibrated spectrophotometric observations from SpeX (gray swathe) and Kast (yellow).}
    \label{fig:ed_fig3}
\end{figure}

\begin{figure}[H]
    \centering
   \includegraphics[width=\textwidth]{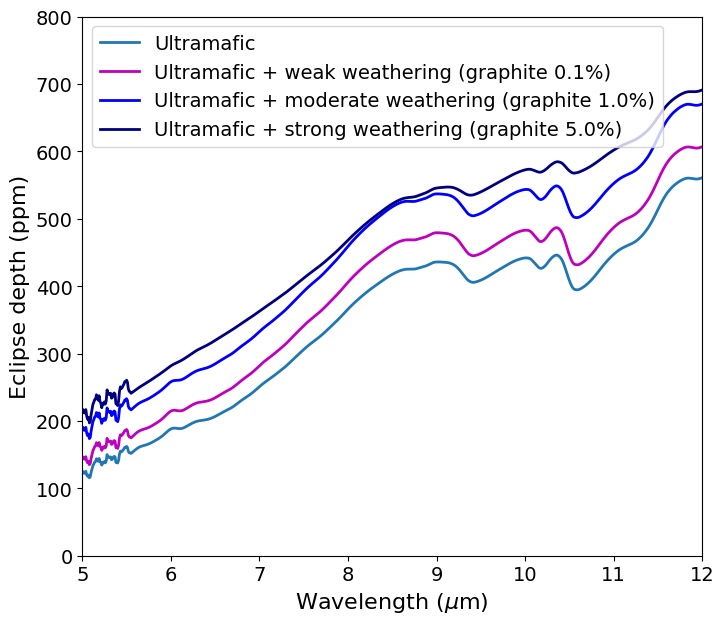}
   \captionsetup{labelformat=empty}
    \caption{\textbf{Extended Data Fig. 4  $\vert$ Effect of space weathering on the emission spectrum of an airless SPECULOOS-3\,b.} Mid-infrared eclipse depths increase with stronger weathering of an ultramafic surface (see Methods for details).}
    \label{fig:ed_fig4}
\end{figure}

 \begin{table}[!htbp]
 \renewcommand\thetable{1}
\begin{tabular}{@{}lll@{}}
\toprule
\textbf{Stellar parameters} &  &  \\
\midrule
Parameter & Value + 1 $\sigma$ error & Unit  \\
\botrule
Spectral type  & M$6.5 \pm0.5$ & \\ 
Parallax & 59.701 $\pm$ 0.043 & mas \\
Distance & $16.750 \pm 0.012$ &  parsec \\
Right ascension & 20:49:27.440 & HH:MM:SS (J2000) \\
Declination & +33:36:50.96& DD:MM:SS (J2000)  \\
Proper motion (RA) & $-207.809 \pm 0.038$ & mas/yr \\
Proper motion (DEC) & $-412.815 \pm 0.037$ & mas/yr  \\
Radial velocity & $17.816 \pm 0.019$ & km\,s$^{-1}$  \\
$U$ galactic velocity & $47.16 \pm 0.29$ & km\,s$^{-1}$  \\
$V$ galactic velocity & $21.34 \pm 0.43$ & km\,s$^{-1}$  \\
$W$ galactic velocity & $-3.68 \pm 0.26$ & km\,s$^{-1}$  \\
$G$ magnitude & $15.380 \pm 0.003$ & mag \\
$K$ magnitude & $10.541 \pm 0.016$ & mag  \\
Mass   & $0.1009 \pm 0.0024$  & $M_\odot$    \\ 
$[Fe/H]$   & $+0.07 \pm 0.10$   & dex   \\
Luminosity & $0.000835 \pm 0.000019$ & $L_\odot$  \\
$T_{eff}$  & $2800 \pm 29$  & K   \\
Radius  & $0.1230 \pm 0.0022$  & $R_\odot$   \\
Density  & $54.7 \pm 2.5$ & $\rho_\odot$  \\
$\log_{10}$ surface gravity (cgs) & $5.265 \pm 0.014$ & dex  \\
$v\sin{i}$ & $4.2 \pm 0.4$ & km\,s$^{-1}$ \\
Age & 6.6$^{+1.8}_{-2.4}$ & Gyr  \\
\toprule
\textbf{Planetary parameters} &  &  \\
\midrule
Orbital period & 0.71912603 $\pm$ 0.00000057 & d \\
Mid-transit time & 2459790.58344 $\pm$ 0.00032 & BJD$_\mathrm{TDB}$ \\
Transit depth $(R_p / R_\ast)^2$  & 5291 $\pm$  116 & ppm \\
Transit impact parameter  & $0.124 \pm 0.085$ &  $R_\ast$ \\
Transit duration & 27.36 $\pm$ 0.23 & min \\
Orbital inclination  & $89.44 \pm 0.39$ & deg \\
Orbital semi-major axis  & 0.007330 $\pm 0.000055$ & au \\
 & 12.81 $\pm$ 0.20& $R_\ast$ \\
Equilibrium temperature $T_{eq}$ & $553\pm 8$ & K \\
(assuming $A_B=0$ and full heat redistribution.) & & \\
Irradiation  & 15.54 $\pm$ 0.42& $S_\oplus$ \\
Radius & 0.977 $\pm$ 0.022 & $R_\oplus$ \\
 & $0.07273 \pm 0.00080$ & $R_\ast$ \\
\botrule
\end{tabular}
\caption{\textbf{Parameters of the SPECULOOS-3 system.} This table presents the properties of the SPECULOOS-3 system as gathered from differents sources or derived through a global Bayesian analysis of the transit photometry (see Methods).}\label{tab1}%
 \label{Table1}
\end{table}

\setcounter{table}{0}
\renewcommand{\thetable}{\arabic{table}} 
\renewcommand{\tablename}{Extended Data Table}

\begin{table*}[!htbp]
\begin{center}
\begin{small}
\begin{tabular}{|c|cc|}
    \hline
    & 10 bar CO$_2$ & 10 bar H$_2$O  \\
    \hline
    Blackbody & \textbf{4} & \textbf{4} \\
    Basaltic & \textbf{7} & \textbf{6} \\
    10 bar CO$_2$ & -- & \textbf{9} \\
    10 bar H$_2$O & \textbf{9} & -- \\
    \hline
\end{tabular}
{\caption{\textbf{Detectability of atmospheres with MIRI/LRS}. Number of occultation observations with MIRI/LRS needed to distinguish the CO$_2$ and H$_2$O atmospheric models from two common airless planet models. These numbers were obtained using wavelength bins of constant resolution $R$=3. }\label{tab:ED_table1}}
\end{small}
\end{center}
\end{table*}

\begin{table*}[!htbp]
\begin{footnotesize}
\begin{tabular}{|c|ccccccc|}
\hline
 & Blackbody & Metal-rich & Ultramafic & Feldspathic & Basaltic & Granitoid & Fe-oxidized \\
\hline
Blackbody & -- & \textbf{$\geq$30} & \textbf{6} & \textbf{3} & \textbf{17} & \textbf{3} & \textbf{21} \\
Metal-rich & \textbf{$\geq$30} & -- & \textbf{9} & \textbf{4} & \textbf{$\geq$30} & \textbf{3} & \textbf{$\geq$30} \\
Ultramafic & \textbf{10} & \textbf{15} & -- & \textbf{10} & \textbf{20} & \textbf{7} & \textbf{18} \\ 
Feldspathic & \textbf{4} & \textbf{5} & \textbf{14} & -- & \textbf{6} & \textbf{$\geq$30} & \textbf{5} \\
Basaltic & \textbf{$\geq$30} & \textbf{$\geq$30} & \textbf{16} & \textbf{5} & -- & \textbf{4} & \textbf{$\geq$30} \\
Granitoid & \textbf{3} & \textbf{5} & \textbf{9} & \textbf{$\geq$30} & \textbf{5} & -- & \textbf{4} \\
Fe-oxidized & \textbf{$\geq$30} & \textbf{$\geq$30} & \textbf{12} & \textbf{5} & \textbf{$\geq$30} & \textbf{4} & -- \\
\hline
\end{tabular}
\caption{\textbf{Detectability of surface mineralogy with MIRI/LRS}. Number of secondary eclipse observations needed with MIRI/LRS to characterise the surface of SPECULOOS-3 b. 
The distinguishability of models was not analyzed beyond 30 eclipses. Ten eclipse observations should allow to distinguish 50\% of competing surface models with 4 $\sigma$ confidence (and 70\% at 2 $\sigma$).}
\label{tab:ED_table2}
\end{footnotesize}
\end{table*}

\newpage
\bibliography{speculoos3}

\end{document}